\documentclass[12pt]{article}
\usepackage{amsfonts,amssymb,amsmath}
\usepackage{epsfig}
\usepackage{float}

\newcommand{\RR}{\mathbb R}
\newcommand{\CC}{\mathbb C}
\newcommand{\NN}{{\mathbb N}}
\newcommand{\ZZ}{{\mathbb Z}}

\newcommand{\DD}{{\cal D}}
\newcommand{\LG}{\mathfrak L}

\renewcommand{\Re}{\mathop{\mathrm{Re}}}
\renewcommand{\Im}{\mathop{\mathrm{Im}}}
\newcommand{\res}{\mathop{\mathrm{res}}}

\newtheorem{theorem}{Theorem}

\newtheorem{remark}{Remark}
\newtheorem{definition}{Definition}

\newcommand{\beq}{\begin{equation}}
\newcommand{\eeq}{\end{equation}}
\newcommand{\ba}{\begin{array}}
\newcommand{\ea}{\end{array}}
\newcommand{\bea}{\begin{eqnarray}}
\newcommand{\eea}{\end{eqnarray}}
\newcommand{\eps}{{\epsilon}}

\usepackage{graphicx}
\usepackage[usenames,dvipsnames]{color}
\usepackage{transparent} 

\DeclareMathAlphabet{\mathpzc}{OT1}{pzc}{m}{it}

\usepackage{ulem}
\usepackage{cancel}

\begin{document}

\begin{center}
{\bf The finite gap method and the periodic NLS Cauchy problem \\ of the anomalous waves, 
  for a finite number of unstable modes
}  
\vskip 10pt
{\it P. G. Grinevich $^{1,3}$ and P. M. Santini $^{2,4}$}

\vskip 10pt

\vskip 10pt

{\it 
$^1$ L.D. Landau Institute for Theoretical Physics, pr. Akademika Semenova 1a, 
Chernogolovka, 142432, Russia,

\smallskip

$^2$ Dipartimento di Fisica, Universit\`a di Roma "La Sapienza", and \\
Istituto Nazionale di Fisica Nucleare (INFN), Sezione di Roma, \\ 
Piazz.le Aldo Moro 2, I-00185 Roma, Italy}

\vskip 10pt

$^{3}$e-mail:  {\tt pgg@landau.ac.ru}\\
$^{4}$e-mail:  {\tt paolo.santini@roma1.infn.it}
\vskip 10pt

{\today}

\end{center}

\begin{abstract}
The focusing Nonlinear Schr\"odinger (NLS) equation is the simplest universal model describing the modulation instability (MI) of quasi monochromatic waves in weakly nonlinear media, and MI is considered the main physical mechanism for the appearence of  anomalous (rogue) waves (AWs) in nature. In this paper we study, using the finite gap method, the NLS Cauchy problem for generic periodic initial perturbations of the unstable background solution of NLS (what we call the Cauchy problem of the AWs), in the case of a finite number $N$ of unstable modes. We show how the finite gap method adapts to this specific Cauchy problem through three basic simplifications, allowing one to construct the solution, at the leading and relevant order, in terms of elementary functions of the initial data. More precisely, we show that, at the leading order, i) the initial data generate a partition of the time axis into a sequence of finite intervals; ii) in each interval $I$ of the partition, only a subset of ${\cal N}(I)\le N$ unstable modes are ``visible'', and iii) the NLS solution is approximated, for $t\in I$, by the ${\cal N}(I)$-soliton solution of Akhmediev type, describing the nonlinear interaction of these visible unstable modes, whose parameters are expressed in terms of the initial data through elementary functions. This result explains the relevance of the $m$-soliton solutions of Akhmediev type, with $m\le N$, in the generic periodic Cauchy problem of the AWs, in the case of a finite number $N$ of unstable modes.
\end{abstract}

\section{Introduction}

The two Nonlinear Schr\"odinger (NLS) equations 
\beq\label{NLS_foc_defoc}
i u_t +u_{xx}+2 \nu |u|^2 u=0, \ \ u=u(x,t)\in\CC, \ \ \nu=\pm 1
\eeq
are the simplest universal models in the description of the propagation of a quasi monochromatic wave in a weakly nonlinear medium; in particular, they are relevant in water waves \cite{Zakharov,AS}, in nonlinear optics \cite{Solli,Bortolozzo,PMContiADelRe}, in Langmuir waves in a plasma \cite{Sulem}, and in the theory of Bose-Einstein condensates \cite{Bludov,Pita}. For instance, in a nonlinear optics interpretation, $u(x,t)$ is the complex amplitude of the electric field and the self-interacting term $2\nu |u(x,t)|^2$ accounts for the nonlinear response of the medium, in a Kerr like regime, proportional to the light intensity; in a quantum mechanical interpretation, $u(x,t)$ is the wave function and $V(x,t)=-2\nu |u(x,t)|^2$ is the self-induced potential proportional to the probability density. If $\nu=1$, the self-interacting term is self-focusing (from which the name ``self-focusing NLS''), and corresponds to a potential well whose depth increases with the density. If $\nu=-1$, the self-interacting term is defocusing (from which the name ``defocusing NLS''), corresponding to a potential barrier whose height increases with the density. It is clear, from these considerations, that the focusing and defocusing regimes correspond to very different evolutions of the same initial data.

In particular, it is well-known that the elementary solution
\beq\label{background0}
a \exp (2i \nu |a|^2t), \ \ a\in \CC \ \mbox{and constant,}
\eeq
of (\ref{NLS_foc_defoc}), describing Stokes waves \cite{Stokes} in a water wave context, a state of constant light intensity in nonlinear optics, and a state of constant boson density in a Bose-Einstein condensate, is stable, in the defocusing case, under the perturbation of waves with arbitrary wave length, while it is unstable, in the focusing case, under the perturbation of waves of sufficiently large wave length \cite{Talanov,BF,Zakharov,ZakharovOstro,Taniuti,Salasnich}; this modulation instability (MI), present only in the focusing case, is considered as the main cause for the formation of anomalous (rogue, extreme, freak) waves (AWs) in nature \cite{HendersonPeregrine,Dysthe,Osborne,KharifPeli1,KharifPeli2,Onorato2}. In this paper the term AW is deliberately used in a vague sense, and by it we just mean order one (or higher order) coherent structures over the unstable background, generated by MI. This paper and our previous works \cite{GS1,GS2,GS3,GS4} are dedicated to a complete analytic understanding of the deterministic aspects of the dynamics of these coherent structures, in the case of a finite number of unstable modes.

Let us point out that, in oceanography, anomalous (or rogue) waves are defined as waves with a sufficiently high amplitude with respect to the average one, and, in this respect, it is known that NLS soliton solutions over the unstable background can reach unusually large amplitudes if their parameters are specially correlated (see, for instance, \cite{CaliniShober2}).  The application of an amplitude criterion to select such anomalously large amplitude waves among finite-gap solutions of NLS over the background, was recently done in \cite{BET}. Our long-term research plan, actually one of the main motivations of the studies made in this paper and in \cite{GS1,GS2,GS3,GS4}, is to understand the deterministic aspects of the theory, valid for a finite number of unstable modes, as the starting point for the study of the statistical aspects of the theory, when a large number of unstable modes are excited, in order to be able, in particular, to describe analytically the probability of the generation in space-time, due to MI, of anomalously large amplitude waves.

The integrable nature \cite{ZakharovShabat} of the focusing NLS equation allows one to construct a large zoo of exact solutions, corresponding to perturbations of the background, by degenerating finite-gap solutions \cite{ItsRybinSall,BBEIM,Krichever2,Krichever3}, when the spectral curve becomes rational, or using classical Darboux transformations \cite{Matveev0}, dressing techniques \cite{ZakharovShabatdress,ZakharovManakov,ZakharovMikha}, and the Hirota method \cite{Hirota0,Hirota}. Among these basic solutions, we mention the Peregrine soliton \cite{Peregrine}, rationally localized  in $x$ and $t$ over the background (\ref{background0}), the so-called Kuznetsov \cite{Kuznetsov} - Kawata - Inoue \cite{KI} - Ma \cite{Ma} soliton, exponentially localized in space over the background and periodic in time, and the solution found by Akhmediev, Eleonskii and Kulagin in \cite{Akhmed0}, periodic in $x$ and exponentially localized in time over the background (\ref{background}), known in the literature as the Akhmediev breather. Elliptic generalizations were constructed in  \cite{Akhmed2}. A more general one soliton solution over the background (\ref{background0}) can be found, f.i., in \cite{ItsRybinSall, ZakharovGelash1}, corresponding to a spectral parameter in general position. These solutions have also been generalized to the case of multi-soliton solutions, describing their nonlinear interaction, see, f.i., \cite{ItsRybinSall,DGKMatv,Akhm6,ZakharovGelash2}.  Finite-genus representations of AWs were constructed in \cite{BET}; elliptic solutions corresponding to special genus 2 curves with symmetries were studied in \cite{Smirnov1} using the method of \cite{BBM}.  We remark that the Peregrine solitons are homoclinic, describing AWs appearing apparently from nowhere and desappearing in the future, while the multisoliton solutions of Akhmediev type are almost homoclinic, returning to the original background up to a multiplicative phase factor. Generalizations of these solutions to the case of integrable multicomponent NLS equations have also been found \cite{BDegaCW,DegaLomb,DegaLombSommacal}. 

Concerning the NLS Cauchy problems in which the initial condition consists of a perturbation of the exact background (\ref{background0}), what we call the Cauchy problem of the AWs, if such a perturbation is localized, then slowly modulated periodic oscillations described by the elliptic solution of (\ref{NLS_foc_defoc}), for $\nu=1$, play a relevant role in the longtime regime \cite{Biondini1,Biondini2}. If the initial perturbation is $x$-periodic, numerical experiments and qualitative considerations indicate that the solutions of NLS exhibit instead time recurrence \cite{Yuen1,Yuen2,Yuen3,Akhmed3,Simaeys,Kuznetsov2}, as well as numerically induced chaos \cite{AblowHerbst,AblowSchobHerbst,AblowHHShober}, in which the almost homoclinic solutions of Akhmediev type seem to play a relevant role \cite{Ercolani,FL,CaliniEMcShober,CaliniShober1,CaliniShober2}. There are reports of experiments in which the Peregrine and the Akhmediev solitons were observed \cite{CHA_observP,KFFMDGA_observP,Yuen3,Tulin,Onor,trillo,PieranFZMAGSCDR}. Their relevance within some classes of localized initial data for NLS, in the small dispersion regime, was shown in \cite{BT,EKT}; see also \cite{GT,TBETB} for the investigation of their relevance in ocean waves and fiber optics.

The proper tool to solve the periodic Cauchy problem for soliton equations is the finite gap method. Its development started in the papers \cite{Novikov,Dubrovin,ItsMatveev,Lax,MKVM}; it was first applied to the NLS equation in \cite{ItsKotlj}, and its generalization to 2+1 equations was first constructed in \cite{Krichever}. In the paper \cite{GS1} and in the present work we apply it to solve the periodic Cauchy problem of the AWs for the focusing NLS equation 
\beq\label{NLS_foc}
i u_t +u_{xx}+2 |u|^2 u=0, \ \ u=u(x,t)\in\CC;
\eeq
i.e., we study the focusing NLS Cauchy problem on the segment $[0,L]$, with periodic boundary conditions, for a generic, smooth, periodic, zero average, small initial perturbation of the background solution
\beq\label{background}
u_0(x,t)=e^{2it};
\eeq
i.e.:
\beq\label{Cauchy}
\ba{l}
u(x,0)=1+\eps v(x), \ \ \ 0<\eps\ll 1,  \\ v(x+L)=v(x),
\ea
\eeq
where
\beq\label{Fourier}
v(x)=\sum\limits_{j\ge 1}^{\infty}\left(c_j e^{i k_j x}+c_{-j} e^{-i k_j x}\right), \ \ k_j=\frac{2\pi}{L}j .
\eeq
We also assume that the period $L$ be generic, i.e., that $L/\pi$ is not an integer.

\begin{remark}
The simplified form (\ref{background}) of the background (\ref{background0}) is obtained choosing $a=1$, without loss of generality, due to the scaling symmetry of NLS. In (\ref{Fourier}) we have also assumed, without loss of generality, due again to the scaling symmetry of NLS, that the perturbation $v(x)$ has zero average
\beq\label{zero_average}
\int\limits_{0}^L v(x)dx =0,
\eeq
implying that the Fourier coefficient $c_0=(1/L)\int\limits_{0}^L v(x)dx$ be zero. 
\end{remark}

It is well-known that a monochromatic perturbation of (\ref{background}) of wave number $k$ is unstable if $|k|<2$; therefore, defining $N\in\NN$ as
\beq\label{def_N}
N=\left\lfloor \frac{L}{\pi}\right\rfloor ,
\eeq
where $\lfloor x \rfloor$,  $x\in\RR$  denotes the largest integer not greater than $x$, 
the first $N$ modes $\{\pm k_j\},~1\le j \le N$, are linearly unstable, since they give rise to exponentially growing and decaying waves of amplitudes $O(\eps e^{\pm \sigma_j t})$, where the growing rates $\sigma_j$ are defined by
\beq\label{def_ampl}
\sigma_j=k_j\sqrt{4-k^2_j}, \ \ 1\le j \le N,
\eeq 
while the remaining modes are linearly stable, since they give rise to small oscillations of amplitude $O(\eps e^{\pm i \omega_j t})$, where 
\beq
\omega_j=k_j\sqrt{k^2_j -4}, \ \ j>N. 
\eeq
It is also convenient to introduce, for the unstable modes, the angles $\phi_j$'s parametrizing them, and defined by
\beq\label{def_angle0}
\phi_j=\arccos(k_j/2)=\arccos\left(\frac{\pi}{L}j\right), \ \ 0<\phi_j<\pi/2 , \ \ 1\le j \le N,
\eeq
implying that
\beq
k_j=2\cos\phi_j, \ \ \sigma_j=2\sin(2\phi_j),  \ \ 1\le j \le N.
\eeq
These well-known facts are summarized in the following formula \cite{GS1,GS2}, valid for $0\le t\le O(1)$:
\beq\label{linear_stage}
\ba{l}
u(x,t)=e^{2it}\Big[1+\sum\limits_{j=1}^N\Big(\frac{\epsilon|\alpha_j |}{\sin 2\phi_j}e^{\sigma_j t+i\phi_j}\cos[k_j (x-x_j)]\\
+ \frac{\epsilon|\beta_j |}{\sin 2\phi_j}e^{-\sigma_j t-i\phi_j}\cos[k_j (x-\tilde{x}_j)]+O(\eps)\mbox{-oscillations} \Big)\Big] +O(\eps^2) ,
\ea
\eeq
where
\beq\label{def_alpha_beta_0}
\ba{l}
\alpha_j=e^{-i\phi_j}\overline{c_j}-e^{i\phi_j}c_{-j}, \ \ \beta_j=e^{i\phi_j}\overline{c_{-j}}-e^{-i\phi_j}c_j, \\
x_j=\frac{\arg(\alpha_j)+\pi/2}{k_j}, \ \ \tilde{x}_j=\frac{-\arg(\beta_j)+\pi/2}{k_j}, \ \ j=1,\dots,N ,
\ea
\eeq
describing the first linear stage of MI, governed by the focusing NLS equation linearized about the solution (\ref{background}): $\delta u_t +\delta u_{xx}+2\exp(4it)\overline{\delta u}+4\delta u=0$.\\
Therefore \\
{\it the initial datum splits into exponentially growing and decaying waves, respectively the $\alpha$- and $\beta$-waves, each one carrying half of the information encoded into the unstable part of the initial datum, plus small oscillations associated with the stable modes, remaining small during the evolution}.

The $j^{th}$ unstable mode becomes $O(1)$ at times of $O(\sigma_j^{-1}|\log\eps|)$; therefore the most unstable modes, the ones appearing first, are the modes with larger growth rate $\sigma_j$. It follows that, at logarithmically large times, one enters the (first) nonlinear stage of MI, the linearized NLS theory cannot be used anymore, and, to describe the evolution, the full integrability machinery of the finite gap method for NLS must be used. 

Once applied to the specific Cauchy problem (\ref{NLS_foc})-(\ref{zero_average}) of the AWs, we found, remarkably, that the finite gap method undergoes the following three basic simplifications, allowing one to construct the solution, at the leading and relevant order, in terms of elementary functions of the initial data (see \cite{GS1} and this paper).

\textbf{Step 1: Finite-gap approximation.} A generic periodic $O(\eps)$ initial perturbation of the unstable background opens infinitely many $O(\epsilon)$ gaps in the spectral problem. They are organized in pairs, and each pair corresponds to the pair $\pm k_j$ of excited modes of the linearized problem. A finite number $N=\lfloor L/\pi\rfloor$ of these modes are unstable, and the remaining ones are stable. Since the stable modes, giving rise to $O(\epsilon)$ oscillations, correspond to $O(\epsilon)$ corrections to the AWs, {\it one can close the corresponding gaps, keeping open only the $2N$ gaps corresponding to the unstable modes} (no matter how small are these gaps, they will cause $O(1)$ effects on the dynamics, due to the instability). Therefore, using this recipe, we go from an infinite-gap theory to its $2N$-gap approximation. Let us point out that this finite gap approximation, specific to the Cauchy problem of AWs, is non standard. Indeed, in the usual finite-gap approximation, one closes gaps smaller than a certain constant, while, in our case, all gaps are small, and the criterion for closing a gap is the stability of the corresponding mode.

\textbf{Step 2. Explicit analytic approximation of the $\theta$-function parameters.} The above finite-gap approximation of the problem implies that the solutions are represented by ratios of $\theta$-functions of genus $2N$. In general, the parameters in the $\theta$-function formulas are complicated transcendental expressions in terms of the Cauchy data. {\it In the special setting we use, good elementary formulas in terms of the initial data can be written, at the leading and relevant order, for all the parameters used in the arguments of the $\theta$-functions}.

\textbf{Step 3. Elementary approximation of the $\theta$-functions.} The Riemann $\theta$-functions are defined as infinite sums of exponentials over all integer points in $\RR^{2N}$. Due to the presence of the small parameter $\epsilon$, at each time it is sufficient to keep summation only over a subset of the $4^{{N}}$ vertices of the elementary hypercube of this multidimensional lattice containing the trajectory point. Therefore, {\it for a generic $t\ge 0$, the infinite sum of exponentials reduces to a finite sum of $4^{{\cal N}(t)}$  exponentials, with $0\le {\cal N}(t)\le N$, whose arguments are given in terms of elementary functions of the Cauchy data. It turns out that this $t$-dependent representation of the solution in terms of elementary functions coincides, at the leading order, with the ${\cal N}(t)$-soliton solution of Akhmediev type}.
\begin{remark}
We remark that the first attempt to apply the finite gap method to solve the NLS Cauchy problem on the segment, for periodic perturbations of the background, was made in \cite{Tracy}; the fact that, in the $\theta$-function representation of the solution, different finite sets of lattice point are relevant in different time intervals was first observed there, but no connection was established between the initial data and the parameters of the $\theta$-function, and no description of the dynamics in terms of elementary functions was given.
\end{remark}

Using the above three simplifying steps, in \cite{GS1} we have studied the Cauchy problem of the AWs (\ref{NLS_foc})-(\ref{Fourier}) in the particular case in which the initial perturbation excites just one of the unstable modes, say $k_n,~1\le n\le N$:
\beq\label{GS1}
u(x,0)=1+\eps(c_n e^{i k_n x}+c_{-n} e^{-i k_n x}), \ \ 1\le n\le N,
\eeq
distinguishing two cases: 1) the case in which only the corresponding unstable gaps are opened by the initial condition (\ref{GS1}), and 2) the case in which  more than one pair of unstable gaps is opened.

  In the first case 1), the solution describes an exact deterministic alternate recurrence of linear and nonlinear stages of MI, and the nonlinear AW stages are described by the Akhmediev breather, whose parameters, different at each AW appearence, are always given in terms of the initial data through elementary functions \cite{GS1}. This result is summarized in the following

  \begin{theorem}

  Consider the case in which the initial condition (\ref{GS1}) opens only the pair of gaps associated with the corresponding modes $\pm k_n$ of the linearized problem, and the associated finite gap Riemann surface is a genus $2N=2$ hyperelliptic curve with two $O(\eps)$ handles. This happens, for instance, if $N=n=1$ ($\pi< L <2 \pi$), the simplest case in which only the mode $k_1=2\pi/L$ is unstable, or if $L/2\pi <n <L/\pi$. 

Introduce the following parameters, for $m\in \NN_+ $:
\beq\label{parameters_recurrence}
\ba{l}
X^{(m)}_n =X^{(1)}_n+(m-1)\Delta X_n, \ \ T^{(m)}_n=T^{(1)}_n+(m-1)\Delta T_n, \\ \Phi^{(m)}_n =2\phi_n+(m-1)4\phi_n , \\
X^{(1)}_n=x_n=\frac{\arg(\alpha_n)+\pi/2}{k_n}, \ \ \Delta X_n=\frac{\arg(\alpha_n\beta_n)}{k_n}, \ \ \mbox{mod }L, \\
T^{(1)}_n= \frac{1}{\sigma_n}\log\left(\frac{\sigma^2_n}{2 \eps |\alpha_n| } \right), \  \ \
\Delta T_n=\frac{2}{\sigma_n}\log\left( \frac{\sigma^2_n}{2\eps \sqrt{|\alpha_n\beta_n |}} \right),
\ea
\eeq
where $\alpha_n$ and $\beta_n$ are defined in (\ref{def_alpha_beta_0}) in terms of the initial data.

Then, for $0\le t\le O(1)$, we are in the first linear stage of MI:
\beq
\ba{l}
u(x,t)=e^{2it}\Big\{1+\frac{1}{\sin(2\phi_n)}\Big[|\alpha_n |\cos\Big(k_n  (x-x_n)\Big)e^{\sigma_n t+i\phi_n}+ \\
|\beta_n |\cos\Big(k_n (x-\tilde{x}_n)\Big)e^{-\sigma_n t-i\phi_n}  \Big] \Big\}+O(\eps^2),
\ea
\eeq
where $x_n,\tilde{x}_n$ are defined in (\ref{def_alpha_beta_0}) 
and, for $|t-T^{(1)}_n|\le O(1)$, we are in the first nonlinear stage of MI, describing the first appearance of the AW through the formula
\beq\label{AW_1st_appearance}
\ba{l}
u(x,t)=e^{2i\phi_n} F\Big(x,t;\phi_n,X^{(1)}_n,T^{(1)}_n \Big)+O(\eps), 
\ea
\eeq
where function $F$  is the Akhmediev breather:
\beq\label{Akhm1a}
\ba{l}
F(x,t;\theta,X,T)\equiv e^{2it}\frac{\cosh[\sigma(\theta) (t-T)+2i\theta ]+\sin\theta \cos[k(\theta)(x-X)]}{\cosh[\sigma(\theta) (t-T)]-\sin\theta \cos[k(\theta)(x-X)]}, \\
~ \\
k(\theta)=2\cos\theta, \ \ \ \ \sigma(\theta)=k(\theta)\sqrt{4-k^2(\theta)}=2\sin(2\theta),
\ea
\eeq
exact solution of focusing NLS for all real values of the parameters $\theta,~X,~T$. The subsequent evolution is completely fixed by the recurrence property of the solution
\beq\label{periodicity}
u(x+\Delta X_{n},t+\Delta T_{n})=e^{2i\Delta T+4i\phi_{n}}u(x,t)+O(\eps^2), \ \ x\in [0,L], \ \ t\ge 0.
\eeq
\end{theorem}

\begin{figure}[H]
\includegraphics[width=14cm]{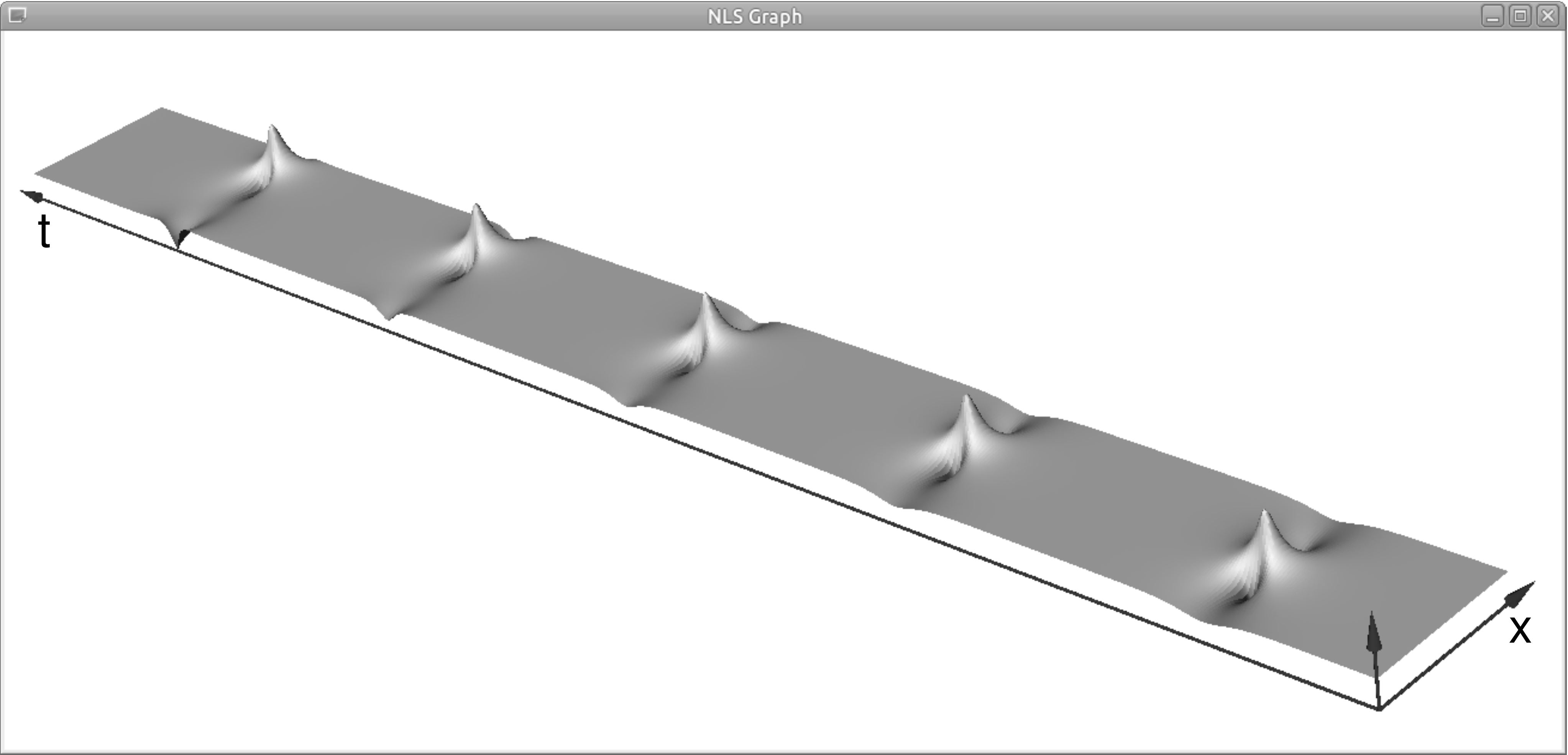}
\caption{\label{fig:1mode} The level plot of $|u(x,t)|$, $x\in [-L/2,L/2]$, describing the alternate appearance of linear and nonlinear stages of modulation instability (the AW sequence), obtained through the numerical integration of NLS via the refined \cite{JR} Split Step Fourier Method \cite{Agrawal,SSFM1,SSFM2}. Here $L=6$ (the case of one unstable mode $k_1$), with $c_1 = 1/2,~c_{-1} = (0.3-0.4 i)/2,  ~\epsilon = 10^{-4}$. The numerical output is in perfect qualitative and quantitative agreement with the theoretical formulas (\ref{parameters_recurrence})-(\ref{AW_sequence}). } 
\end{figure}

\begin{remark}
  We first observe that the first linear and nonlinear stages of MI do match in the intermediate region $O(1)\ll t\ll T^{(1)}_n=O(\sigma^{-1}_n|\log\eps |)$. We also remark that the periodicity property (\ref{periodicity}) implies that the solution describes an exact recurrence of AWs, and the $m^{th}$ AW of the sequence ($m\ge 1$) is described, in the time interval $|t-T^{(m)}_n|\le O(1)$, by the analytic deterministic formula
\beq\label{AW_sequence}
\ba{l}
u(x,t)=e^{i\Phi^{(m)}_n} F\Big(x,t;\phi_n,X^{(m)}_n,T^{(m)}_n \Big)+O(\eps), \ \ m\ge 1 .
\ea
\eeq 
\end{remark}
Therefore we have the following simple picture.\\
{\it The solution of the $x$-periodic Cauchy problem (\ref{NLS_foc})-(\ref{zero_average}) describes, in the case in which the initial condition (\ref{GS1}) opens only the pair of gaps associated with $\pm k_n$, an exact recurrence of Akhmediev breathers, whose parameters, changing at each appearance, are expressed in terms of the initial data via elementary functions. $T^{(1)}_n$ is the first appearance time of the AW (the time at which the AW achieves the maximum of its modulus), $X^{(1)}_n+Lj/n$, $1\le j\le n-1$ are the positions of such a maximum, $1+2\sin\phi_n$ is the value of the maximum, $\Delta T_n$ is the recurrence time (the time interval between two consecutive AW appearances), $\Delta X_n$ is the $x$-shift of the position of the maxima in the recurrence. In addition, after each appearance, the AW changes the background by the multiplicative phase factor $\exp(4i\phi_n)$ (see Figure~\ref{fig:1mode}). }
\begin{remark} {\bf On the physical relevance of the above exact recurrence of AWs} It is very important to remark that, if the number of unstable modes is greater than one ($N>1$), this uniform in $t$ dynamics is sensibly affected by perturbations due to numerics and/or real experiments. Indeed these perturbations open generically other small unstable gaps, provoking $O(1)$ corrections to the result. Therefore this analytic and uniform in $t$ result is physically relevant (and numerically verifiable) only when $N=n=1$ ($\pi < L < 2\pi$).
\end{remark}

2) In the second case, in which more than one unstable gap is opened by the initial condition (\ref{GS1}), a detailed investigation of all these gaps is necessary to get a uniform in $t$ dynamics, and this study is postponed to a subsequent paper. It was however possible to obtain in \cite{GS1} the elementary description of the first nonlinear stage of MI, described again by the Akhmediev breather solution, and how perturbations due to numerics and/or real experiments can affect this result.

\begin{remark}
{\bf AW recurrence as basic effect of nonlinear MI in the periodic setting, and the finite gap method} The recurrence of AWs can be predicted from simple qualitative considerations. The unstable mode grows exponentially and becomes $O(1)$ at logarithmically large times, when one enters the nonlinear stage of MI, and one expects the generation of a transient, $O(1)$, coherent structure over the unstable background (the AW). Since the Akhmediev breather describes the one-mode nonlinear instability, it is the natural candidate to describe such a stage, at the leading order. Due again to MI, this AW is expected to be destroyed in a finite time interval, and one enters the third asymptotic stage, characterized, like the first one, by the background plus an $O(\eps)$ perturbation. This second linear stage is expected, due again to MI, to give rise to the formation of a second AW (the second nonlinear stage of MI), described again by the Akhmediev breather, but, in general, with different parameters. And this procedure iterates forever, in the integrable NLS model, giving rise to the generation of an infinite sequence of AWs described by different Akhmediev breathers. {\it Then the AW recurrence is a relevant effect of nonlinear MI in the periodic setting, and the finite gap method is the proper tool to give an analytic description of it}. 
\end{remark}
\begin{remark}
{\bf Finite gap approach vs matched asymptotic expansions.} The above AW recurrence is described by an alternating sequence of linear and nonlinear asymptotic stages of MI, obviously matching in their overlapping time regions; therefore this finite gap result naturally motivates the introduction of a matched asymptotic expansions (MAEs) approach, presented in the paper \cite{GS2}, and involving more elementary mathematical tools. The advantages of the finite gap approach are due to the fact that the $\theta$-function representation of the solution is uniform in time, and the analytic description of the nonlinear stages of MI (of the sequence of AWs) does not require any guess work. Such a guess work is instead needed in the MAE approach, when one has to select the proper nonlinear coherent structure of NLS describing a certain nonlinear stage of MI and matching with the preceeding linear stage. In all the situations in which such a guess work is no problem, the MAE approach becomes competitive, since it involves more elementary mathematics, like in the case of one unstable mode only. But, as we shall see in the following sections, if we have more than one unstable mode, and the initial data are generic, the dynamics is not described by a sequence of asymptotic stages of MI, and the MAE approach is not applicable, while the finite gap method is able to provide the uniform representation of the solution. The MAE approach works, in the case of a finite number $N>1$ of unstable modes, only for very special initial data \cite{GS2}.   
\end{remark}  
\begin{remark}
{\bf AW recurrence as stable output of the dynamics. } Theorem 1 explains the relevance of the Akhmediev breather in a generic periodic Cauchy problem of AWs, in the case of one unstable mode, and leads to the following natural question: what happens if the initial condition is the highly non generic Akhmediev breather? The answer is the following. The Akhmediev breather, quasi homoclinic solution of NLS, is unstable, corresponding, in the finite gap spectral picture, to the highly non generic case in which all the spectral gaps are closed. Any small perturbation (due, for instance, to the numerical scheme approximating NLS, or to small corrections to NLS coming from physics) opens small gaps, implying that, after the first appearance of the AW as predicted by the initial datum, a recurrence of AWs, described by (\ref{parameters_recurrence})-(\ref{AW_sequence}), will be the stable output of the dynamics \cite{GS3}.
\end{remark}
\begin{remark}
{\bf AW recurrence for other NLS type models.} It is natural to ask if the above AW recurrence is typical of NLS, or it is shared by other integrable NLS type models. In the case of the focusing Ablowitz - Ladik model \cite{AblowLadik} $i {u_n}_t+u_{n+1}+u_{n-1}-2 u_n +|u_n|^2(u_{n+1}+u_{n-1})=0,~n\in\ZZ$, integrable discretization of focusing NLS, the picture is essentially the same. Indeed, using MAEs and in the case of one unstable mode only, it was shown in \cite{Coppini} that a generic periodic initial datum leads to an AW recurrence of Narita solutions (the Narita solution is the discrete analogue of the Akhmediev breather \cite{Narita,Akhmed7}). In the case of the PT-symmetric NLS (PT-NLS) equation \cite{AM1,AM2,AM3,AM4} $i w_t(x,t) +w_{xx}(x,t)+2 w^2(x,t)\overline{w(-x,t)}=0, \ w=w(x,t)\in\CC$, there are instead novelties with respect to the NLS case. The analogue of the Akhmediev breather, always regular for all values of its parameters, is now a pair of exact solutions of PT-NLS \cite{san1}. The first one is singular in a certain domain of its parameter space, while the second one is always singular, for all values of its arbitrary parameters; these singularities describe blow ups in points of the $(x,t)$ plane. It follows that, depending on the initial data, the AWs are either regular, with arbitrarily large amplitude, or they blow up at finite time \cite{san2}. This is a novel phenomenon for integrable NLS type systems, and a qualitative reason for it should be the gain-loss properties of the complex self-induced potential of PT-NLS, causing extra-focusing effects with respect to the NLS case.    
\end{remark}
\begin{remark}
  {\bf NLS exact recurrence vs Fermi-Pasta-Ulam recurrence in nature} From \cite{GS1,GS2} and formulas (\ref{parameters_recurrence})-(\ref{AW_sequence}) it follows that, if the initial condition (\ref{Cauchy}) excites the only unstable mode $k_1$: $u(x,0)=1+\eps(c_1\exp(ik_1 x)+c_{-1}\exp(-ik_1 x))$, then: i) the energy is initially concentrated on the zero mode (the background) and on the first mode (the monochromatic perturbation):
  \beq
|u_0(0)|^2=1, \ \ |u_m(0)|^2=\delta_{m,\pm 1}\eps^2 |c_{\pm 1}|^2,
  \eeq
where $u_m(t),~m\in\ZZ$ are the Fourier coefficients of the NLS solution $u(x,t)$. ii) At the first appearance time $T^{(1)}_1$ of the AW (see (\ref{parameters_recurrence})), the energy is distributed on all Fourier modes according to the simple law
\beq
\ba{l}
|u_0(T^{(1)}_1)|^2=(2\cos\phi_1 -1)^2, \ \ 0<\phi_1 <\frac{\pi}{2}, \\ |u_m(T^{(1)}_1)|^2=4(\cos\phi_1)^2\left(\tan\left(\frac{\phi_1}{2} \right) \right)^{2|m|}, \ \ m\ne 0,
\ea
\eeq
  but, iii) at the recurrence time $\Delta T_1$, it is re-absorbed by the zero and first modes:
  \beq
|u_0(\Delta T_1)|^2=1, \ \ \ |u_m(\Delta T_1)|^2=\delta_{m,\pm 1}\eps^2 |c_{\pm 1}|^2,
  \eeq
  starting an exact recurrence \cite{GS5}.
  
  AW recurrence in the periodic setting has been already considered in the literature (see, f.i., \cite{Yuen1,Yuen2,Yuen3,Akhmed3,Simaeys,Kuznetsov2}), and recent experiments in water waves \cite{Onor}, in fiber optics \cite{trillo}, and in the nonlinear optics of a photorefractive crystal \cite{PieranFZMAGSCDR} accurately reproduce the recurrence phenomena. In particular, the experimental findings obtained in \cite{PieranFZMAGSCDR} have been compared with the formulas (\ref{parameters_recurrence})-(\ref{AW_sequence}) describing the NLS recurrence, obtaining a very good qualitative and quantitative agreement. 1) Since NLS describes the above different physics only at the leading order, one expects that the exact NLS recurrence of AWs, illustrated in formulas (\ref{parameters_recurrence})-(\ref{AW_sequence}), be replaced by a ``Fermi-Pasta-Ulam'' - type recurrence \cite{Fermi_Pasta_Ulam}, before thermalization destroys the pattern. Indeed, in \cite{PieranFZMAGSCDR}, up to 3 recurrences were observed and compared with the above NLS recurrence formulas, obtaining a very good qualitative and quantitative agreement. It was also shown that the recurrent behavior disappears when the photorefractive crystal works, instead, in a nonlinear regime different from the integrable (Kerr) one. 2) A common feature of the experiments \cite{Onor,trillo,PieranFZMAGSCDR} is the choice to work, in most of the cases, in a special symmetry of the experimental apparatus leading to the particularly significant subcases in which the recurrence shift $\Delta X$ in (\ref{parameters_recurrence}) is either $0$ or $L/2$, implying respectively the time-periods $\Delta T_1$ or $2 \Delta T_1$, where $\Delta T_1$ is the recurrence time (\ref{parameters_recurrence}). This particular symmetry corresponds to the distinguished sub-case in which $|c_1|\sim |c_{-1}|$ in (\ref{Fourier}), and leads to interesting physical resonances of the physical times $T^{(1)}_1$ and $\Delta T_1$ as functions of $\zeta=\frac{\arg c_1 +\arg c_{-1}}{2}$, for $\zeta=-\phi_1,~\pi-\phi_1$ \cite{GS4}; these resonances have been experimentally observed in \cite{PieranFZMAGSCDR}. The above experimental findings, in good quantitative agreement with the theoretical formulas (\ref{parameters_recurrence})-(\ref{AW_sequence}), are an important confirmation that NLS is a good model in the description of nonlinear modulation instabilities in nonlinear optics and water waves. 
\end{remark}

In this paper we apply the finite gap method to the generic periodic Cauchy problem (\ref{NLS_foc})-(\ref{zero_average}) of the AWs, in the case of a finite number $N$ of unstable modes. Qualitative considerations similar to those made in Remark 5 suggest again that each unstable mode will appear recurrently in the dynamics, depending on its degree of instability; but this recurrence now is affected by the nonlinear interactions with all the other unstable modes (we shall see that, at the leading order, this interaction is pairwise). Again the proper tool to describe all that is the finite gap method, and again the finite gap method adapts to this specific Cauchy problem through the three basic simplifications outlined above, allowing one to construct the solution, at the leading and relevant order, in terms of elementary functions of the initial data also in this more complicated case.

More precisely, we will show that, at the leading order, i) the initial data generate a partition of the time axis into a sequence of finite intervals; ii) in each interval $I$ of the partition, only a subset of ${\cal N}(I)\le N$ unstable modes are ``visible'', and iii) the NLS solution is approximated, for $t\in I$, by the ${\cal N}(I)$-soliton solution of Akhmediev type, describing the nonlinear interaction of these visible unstable modes, whose parameters are expressed in terms of the initial data through elementary functions. This result explains the relevance of the $m$-soliton solutions of Akhmediev type, with $m\le N$, in the generic periodic Cauchy problem of the AWs, and in the case of a finite number $N$ of unstable modes. Therefore, {\it in the case of a finite number of unstable modes, the theory of NLS anomalous waves is completely deterministic, and its analytic description, at the leading and relevant order, is given in terms of elementary functions of the Cauchy data}.

The paper is organized as follows. In Section \ref{sec:sec2} we describe the main results. In Section \ref{sec:periodicNLS}  we summarize the classical features of the periodic Cauchy problem for the focusing and defocusing NLS equations. In Section \ref{sec:sec4}  we apply this theory to the Cauchy problem of the anomalous waves, constructing the main and auxiliary spectra at the leading and relevant order, through elementary functions of the initial data. In Section \ref{sec:sec5}, after closing the infinitely many gaps corresponding to the stable modes, obtaining a non standard finite gap approximation (the first basic simplification of the theory), we study the corresponding finite gap curve, constructing the leading order expression of the Riemann matrix, of the vector of Riemann constants, and of all the other quantities appearing in the parameters of the $\theta$-functional formulas of the inverse problem, in terms of elementary functions (this is the second basic simplification of the theory). In Section \ref{sec:sec6} we write the $\theta$-functional formulas for the leading order solution, and Section \ref{sec:sec7} is devoted to the presentation of the third basic simplification of this theory, in which the infinite sum of exponentials appearing in the definition of the $\theta$ function, is reduced to a sum over a finite subset of exponentials, different in different time intervals. 
\vskip 10pt
\noindent
This paper will appear in Novikov's volume, on the occasion of his 80th birthday. In this respect, it is worth mentioning that

1) Novikov has always been interested in applying his results to real physics.

2) Novikov has always pointed out that the finite-gap formulas require additional
effectivization to become applicable.

We hope we made here a serious progress in both directions. This paper is dedicated to S.P. Novikov.

\section{Results}
\label{sec:sec2}

The aim of this paper is to provide the solution, at the leading order and in terms of elementary functions, of the generic periodic Cauchy problem of the AWs (\ref{NLS_foc})-(\ref{zero_average}) described in the Introduction, rewritten here for completeness:
\beq\label{Cauchy_AWs}
\ba{l}
  i u_t +u_{xx}+2 |u|^2 u=0, \ \ u=u(x,t)\in\CC , \ \ x\in [0,L], \ \ t\ge 0,\\
  u(x+L,t)=u(x,t),\\
u(x,0) = 1 + \epsilon v(x), \ \ |\epsilon|\ll 1, \\
v(x)=\sum\limits_{j\ge 1}\left(c_j e^{i k_j x}+c_{-j} e^{-i k_j x}\right), \ \ k_j=\frac{2\pi}{L}j , \ \ |c_j |=O(1),
\ea
\eeq
where the period $L$ is generic ($L/\pi$ is not an integer).

  We first list the ingredients we need to construct its solution. \\
1. The number $N$ of the unstable modes
\beq\label{eq:generic2}
N = \left\lfloor \frac{L}{\pi} \right\rfloor ;
\end{equation}
2. their wave numbers $k_j$ and growth rates $\sigma_j$:
\beq\label{eq:wave_numbers}
\ba{l}
k_j=\frac{2\pi}{L}j, \ \ \ \ \ \sigma_j=k_j\sqrt{4-k_j^2}, \ \ \ \ \ 1\le j \le N, \\
\ea
\end{equation}
and the angles $\phi_j$ parametrizing them:
\beq
\phi_j=\arccos\left(\frac{k_j}{2}j \right)=\arccos\left(\frac{\pi}{L}j \right), \ \ 0 < \phi_j < \frac{\pi}{2}, \ \ 1\le j \le N,
\eeq
so that
\beq
k_j = 2 \cos(\phi_j), \ \ \ \ \sigma_j = 2 \sin(2\phi_j).
\eeq
We also define the $2N$ angles $\hat\phi_j$ by:
\begin{equation}
\label{eq:anglesb}
\hat\phi_j=\phi_j, \ \ \hat\phi_{j+N}=-\phi_j, \ \ j=1,\ldots,N.
\end{equation}
3. The following linear combinations of the Fourier coefficients of the unstable modes:
\beq\label{def_alpha_beta}
\alpha_j= (e^{-i\phi_j} \overline{c_j}-e^{i\phi_j}c_{-j}), \ \beta_j= (e^{i\phi_j} \overline{c_{-j}}-e^{-i\phi_j}c_{j}), \ \ j=1,\ldots,N ;
\eeq
and the quantities
\beq\label{def_hat_alpha_beta}
\hat\alpha_j=\alpha_j, \ \ \hat\alpha_{j+N}=\overline{\beta_j},  \ \  \hat\beta_j=\beta_j, \ \ \beta_{j+N} = \overline{\alpha_j}, \ \ j=1,\ldots,N .
\eeq
4. The leading order $2N\times 2N$ Riemann matrix $B=(b_{jk})$:
\begin{equation}
\label{eq:riemann1}
b_{jj}= 2 \log\left[ \frac{\epsilon \sqrt{\hat\alpha_j\hat\beta_j}}{|4\sin(2\hat\phi_j)\cos(\hat\phi_j) |} \right], \ \ j=1,\ldots,2N,
\end{equation}

\begin{equation}
\label{eq:b_off_diag_a}
b_{jk}= 2 \log \left|\frac{\sin\left(\frac{\hat\phi_j-\hat\phi_k}{2} \right)}{\cos\left(\frac{\hat\phi_j+\hat\phi_k}{2} \right) }\right|, \ \ j\ne k, \ \ j,k=1,\ldots,2N,
\end{equation}
assuming that
$$
\Re \sqrt{\hat\alpha_j\hat\beta_j}\ge 0, \ \  j=1,\ldots,2N ,
$$
$$
\sqrt{\hat\alpha_{j+N}\hat\beta_{j+N}} =\overline{\sqrt{\hat\alpha_{j}\hat\beta_{j} }}, \ \  j=1,\ldots,N.
$$
5. The following $2N$ dimensional \textbf{complex} vectors:
\begin{equation}
\label{eq:z_minus}
\big(\vec z_{-}(x,t)\big)_j= \frac{i\pi}{2}  - \log\left[\frac{\hat\alpha_j}{\sqrt{\hat\alpha_j\hat\beta_j}} \right]  + 2i\cos(\hat\phi_j) x -2\sin(2\hat\phi_j) t,
\end{equation}
$$
\big(\vec z_{+}(x,t)\big)_j = \big(\vec z_{-}(x,t)\big)_j -\pi i - 2 i \hat\phi_j ,  \ \  j=1,\ldots,2N,
$$
describing the leading order linearized NLS evolution on the Jacobian variety. We see that 
\begin{equation}
\label{eq:w}
\ba{l}
\Re (\vec z_{+}(x,t))_j = \Re (\vec z_{-}(x,t))_j  =  -2\sin(2\hat\phi_j)t - \frac{1}{2} \log\left|\frac{\hat\alpha_j}{\hat\beta_j}  \right|,\\
\Re (\vec z_{\pm}(x,t))_{j+N}=-\Re (\vec z_{\pm}(x,t))_{j}, \ \ j=1,\dots,N ,
\ea
\end{equation}
and these expressions do not depend on $x$.\\
6. The $2N$-dimensional \textbf{real} vector:
\begin{equation}
\label{eq:vec_w}
\vec w(t) = (\Re B)^{-1} \Re \vec z_{-}(x,t) =  (\Re B)^{-1} \Re \vec z_{+}(x,t),  
\end{equation}
describing the corresponding straight line time evolution in $\RR^{2N}$, equipped with the standard integer lattice $\ZZ^{2N}$.
From (\ref{def_alpha_beta}), (\ref{eq:b_off_diag_a}), (\ref{eq:w}), it follows immediately that
\begin{equation}
\label{eq:vec_w2}
w_{j+N}(t) \equiv -w_j(t), \ \ j=1,\ldots,N.  
\end{equation}
Equation (\ref{eq:vec_w2}) suggests that the construction of the relevant components of $\vec w(t)$ can be simplified, involving the inversion of a simpler to handle $N\times N$ matrix. Indeed, if we introduce the N-vector  $\underline{w}(t)\in\RR^N$ whose components are the first $N$ components of $\vec w(t)$, then:
\beq\label{def_N_vec_w}
\ba{l}
\vec w(t)=(\underline{w}(t),-\underline{w}(t))^T\in \RR^{2N}, \\
\underline{w}(t)= \underline{w}^{(1)} t+\underline{w}^{(0)}, \\
\underline{w}^{(1)}=-{\cal B}^{-1}\underline{\sigma} , \ \ \ \ \underline{w}^{(0)}=-{\cal B}^{-1}\underline{\chi} ,
\ea
\eeq
where ${\cal B}^{-1}$ is the inverse of the real symmetric $N\times N$ matrix ${\cal B}$ defined by
\beq
{\cal B}_{jj}=-\sigma_j \tau_j, \ \  {\cal B}_{jk}= 2\log\left| \frac{\sin(\phi_j -\phi_k)}{\sin(\phi_j +\phi_k)} \right|, \ \ 1\le j,k\le N, \ j\ne k,
\eeq
with
\beq
\tau_j=\frac{2}{\sigma_j}\log\left(\frac{\sigma_j^2}{2\eps\sqrt{\left|\alpha_j \beta_j \right|}} \right), \ \ \ \ 1\le j\le N,
\eeq
and the $N$-vectors $\underline{\sigma}, ~\underline{\chi}\in\RR^N$ are defined by
\beq\label{def_vec_sigma}
\underline{\sigma}=\left(
\ba{c}
\sigma_1 \\
\vdots \\
\sigma_N
\ea
\right), \ \ \ \ \ \underline{\chi}=\left(
\ba{c}
\log(\sqrt{|\alpha_1/\beta_1|}) \\
\vdots \\
\log(\sqrt{|\alpha_N/\beta_N|})
\ea
\right).
\eeq
7. The real number $p$, $0<p\le 1$, characterizing the accuracy of the approximation $\hat u(x,t)$ we want to achieve in the construction of the solution $u(x,t)$: $u(x,t)=\hat u(x,t)+O(\eps^p)$. It follows that, if, at a given time, the contribution of a certain unstable mode to the solution is less than $O(\eps^p)$, this mode is neglected, because invisible with respect to the above approximation.

\begin{figure}[H]
\begin{center}
    \includegraphics[width=10cm]{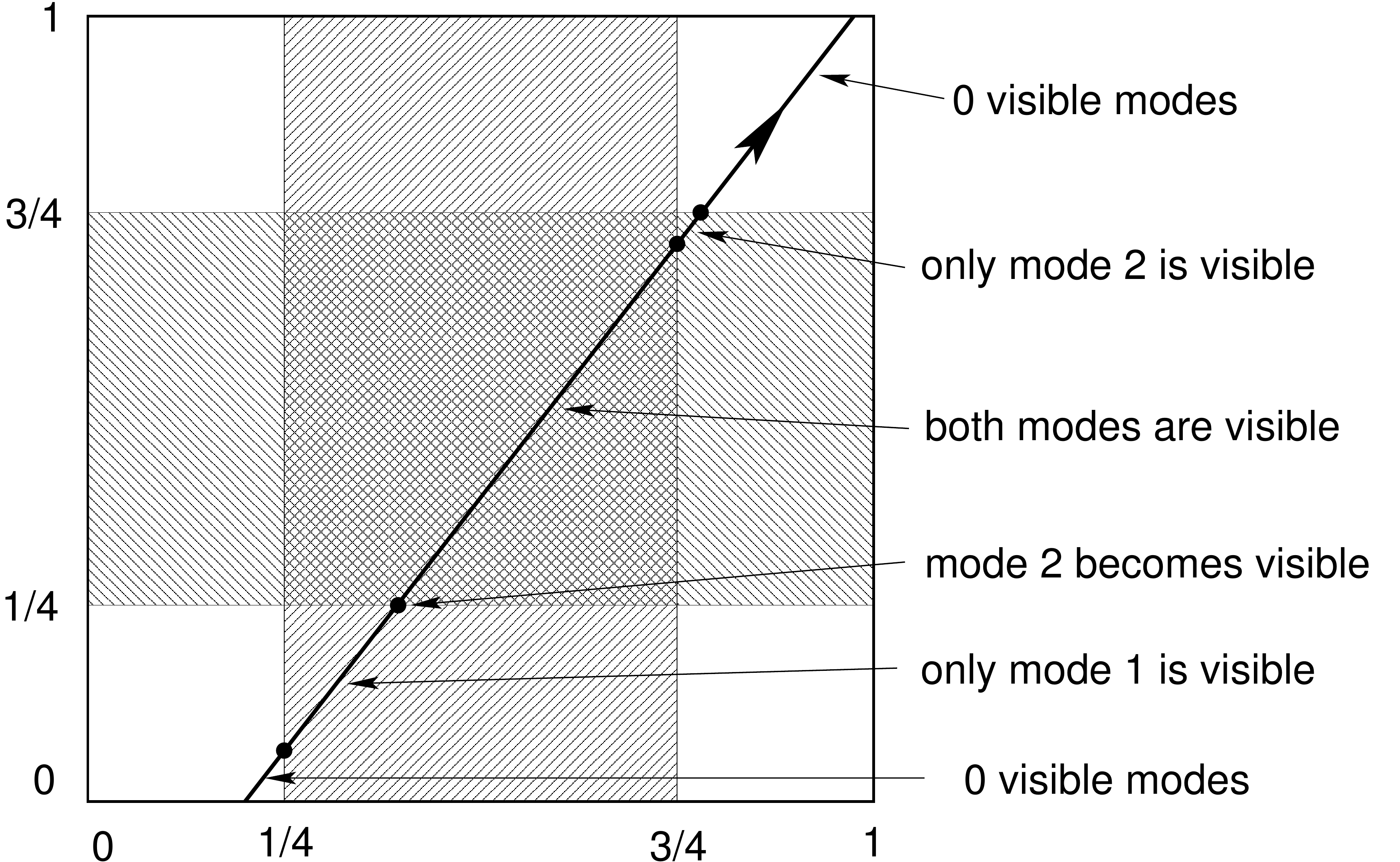}
\end{center}
    \caption{\label{fig:tr} The straight line trajectory $\underline{w}(t)\in \RR^N$ for $N=2$, with $p=1/2$.}

\end{figure}

\begin{definition}
An unstable mode is ``$p$-visible'' at a certain time $t$, if its contribution to the solution is of order $\epsilon^p$, or greater. It is  ``$p$-invisible'' otherwise. 
\end{definition}
8. The following $2N$ vector $\vec{st}(t)\in\RR^{2N}$ of components
\begin{equation}
\label{eq:conters2}
st_j(t) =1- \Theta\left(-w_j(t) +\lfloor w_j(t) \rfloor + \frac{1-p}{2}\right)+\Theta\left(w_j(t) -\lfloor w_j(t) \rfloor -\frac{1+p}{2}\right)    , \ \ j=1,\ldots, 2N,
\end{equation}
where $\Theta(x)$ in (\ref{eq:conters2}) denotes the standard step function:
$$
\Theta(x)=\left\{\begin{array}{ll} 0, & x\le 0, \\ 1, & x>0.  \end{array} \right.
$$
Equivalently,
$$
st_j(t)=\left\{\begin{array}{ll} 
0 & \mbox{if} \ \ w_j(t) - \lfloor  w_j(t) \rfloor < \frac{1-p}{2}, \\
1 & \mbox{if} \ \ \frac{1-p}{2} \le w_j(t) - \lfloor  w_j(t) \rfloor \le \frac{1+p}{2}, \\
2 & \mbox{if} \ \ w_j(t) - \lfloor  w_j(t) \rfloor > \frac{1+p}{2}.  \end{array}\right.
$$
with:
$$
st_{j+N}(t)+st_j(t) = 2, \ \ \mbox{for all} \ \ t \ \ \mbox{and} \ \ j=1,\ldots,N.
$$
The vector (\ref{eq:conters2}) is used to detect which unstable mode is ``$p$-visible'' at a given time. The $j$-th mode is ``$p$-visible'' at time $t$, iff $st_j(t)=1$, and ``$p$-invisible'' if $st_j(t)=0$ or $st_j(t)=2$ (see Figure~\ref{fig:tr}). \\

After the introduction of the above ingredients, all expressed in terms of the initial data via elementary functions, we are ready to formulate the main result, proven in the next Sections.

\begin{figure}[H]
    \includegraphics[width=14cm]{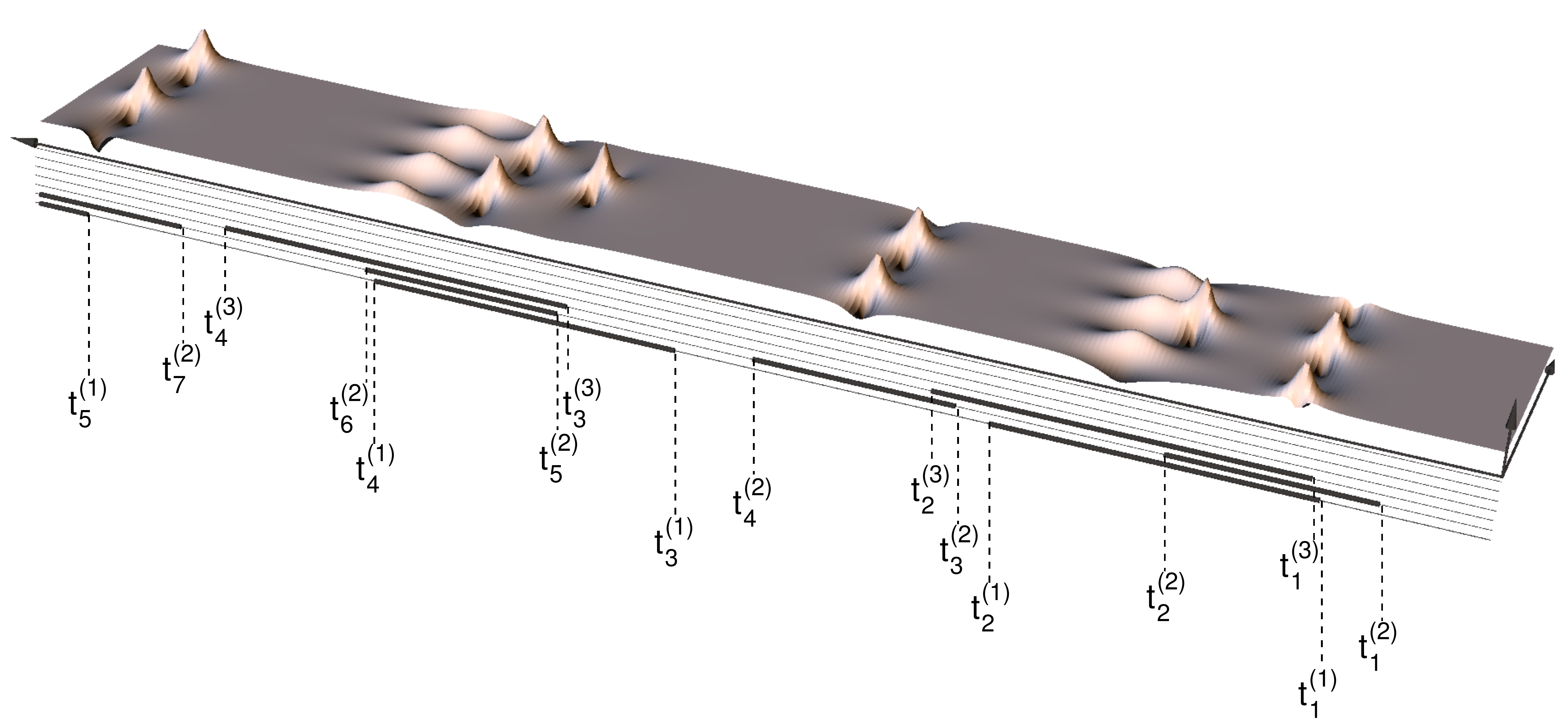}
    \caption{\label{fig:3modes}  The graph of $|u(x,t)|$ at the leading order, from formulas (\ref{sol_approx})-(\ref{phases}), in good agreement with the corresponding numerical experiment. Here $L=10$ (N=3), $0\le t\le 60$, $x\in[-L/2,L/2], $ $\epsilon=10^{-6}$, $c_1 = 0.5$, $c_{-1} = 0.3+0.3i$, $c_2 = 0.5$, $c_{-2} = -0.03+0.03i$, $c_3 = 0.03$, $c_{-3} = 0.02+0.03i$, $p=1/2$, and the short axis is the $x$-axis. The intervals of ``p-visibility'' for the 3 unstable 
modes are marked, below the graph, by bold lines. The boundary points of the partition intervals are, sequentially, $0$, $t^{(2)}_1$, $t^{(1)}_1$, $t^{(3)}_1$, $t^{(2)}_2$, $t^{(1)}_2$, $t^{(2)}_3$, $t^{(3)}_2$,
 $t^{(2)}_4$, $t^{(1)}_3$, $t^{(3)}_3$, $t^{(2)}_5$, $t^{(1)}_4$, $t^{(2)}_6$,  $t^{(3)}_4$, $t^{(2)}_7$, $t^{(1)}_5$. In the first interval ($0,t^{(2)}_1$) we are in the first linear stage of MI, and all modes are invisible; in the interval ($t^{(2)}_1,t^{(1)}_1$) the most unstable mode $2$ is visible; in the interval ($t^{(1)}_1,t^{(3)}_1$) also the mode 1 is visible; in the interval ($t^{(3)}_1,t^{(2)}_2$) all the three modes are visible; and so on. The unstable mode $2$, characterized by two maxima in the period, is the most unstable mode ($\sigma_2=\max\{\sigma_1,\sigma_2,\sigma_3 \}$). It first appears in the interval ($t^{(2)}_1,t^{(2)}_2$); it desappears in the interval ($t^{(2)}_2,t^{(2)}_3$), reappearing again in the interval ($t^{(2)}_3,t^{(2)}_4$); and so on. Its recurrence properties are ruled by equations (\ref{eq:Nrecurrence}). }
\end{figure}

\begin{theorem} Consider the generic periodic Cauchy problem of the anomalous waves (\ref{Cauchy_AWs}), for $x\in [0,L]$ and $t$ in any finite time interval $[0,T_0]$. Introduce the positive quantities $t^{(j)}_{k}$, $j=1,\ldots,N$, $k\ge 1$ defined by
\begin{equation}
\label{eq:change_times} 
t^{(j)}_{2n-1}= \frac{\frac{1-p}{2} + n-1 - w^{(0)}_j }{w^{(1)}_j }, \ t^{(j)}_{2n}= \frac{\frac{1+p}{2} + n - 1 - w^{(0)}_j }{w^{(1)}_j }, \ n\ge 1, \ 1\le j \le N,
\end{equation}
where $w^{(0)}_j,\ w^{(1)}_j, \ 1\le j \le N$, are the $N$ components of vectors $\underline{w}^{(0)},\ \underline{w}^{(1)}$, defined in (\ref{def_N_vec_w})-(\ref{def_vec_sigma}). They are the times at which the $j^{th}$ unstable mode changes its status, from p-invisible to p-visible if $k$ is odd, and from p-visible to p-invisible if $k$ is even, exhibiting the following recurrence properties 
\begin{align}
\label{eq:Nrecurrence}
t^{(j)}_{2n}-t^{(j)}_{2n-1} &= \frac{p}{w^{(1)}_j}, \ \ t^{(j)}_{2n+1}-t^{(j)}_{2n} = \frac{1-p}{w^{(1)}_j},\\
t^{(j)}_{2n+1}-t^{(j)}_{2n-1} &=  t^{(j)}_{2n+2}-t^{(j)}_{2n} = \frac{1}{w^{(1)}_j}, \ \ j=1,\ldots,N, \ \ n\ge 1.\nonumber
\end{align}
They naturally partition the interval $[0,T_0]$ in a sequence of finite intervals.
In a given interval $I$ of the partition, the  $j^{th}$ unstable mode is p-visible iff
\beq
\frac{1-p}{2} \le w_j(t) - \lfloor  w_j(t) \rfloor \le \frac{1+p}{2}, \ \ 1\le j \le N, \ \ t\in I,
\eeq
and p-invisible otherwise. Let us denote by ${\cal N}(I)$ the number of ``$p$-visible'' modes in the interval $I$, and by $s_k(I)$, $k=1,\ldots,{\cal N}(I)$ the indices of these ``$p$-visible'' modes, assuming that $s_{k+{\cal N}(I)}(I)= s_k(I)+N$.  
Then the solution of the Cauchy problem, at the leading and relevant order, is described by the following formula 
\beq\label{sol_approx}
u(x,t) = u_I(x,t) + O(\epsilon^p),  \ \ t\in I,   
\eeq
where $u_I(x,t)$ is the exact ${\cal N}(I)$-soliton solution of Akhmediev type describing the nonlinear interaction of the ${\cal N}(I)$ unstable modes that are visible in the interval $I$, $L$-periodic in $x$ and localized in time over the background:
$$
u_I(x,t)=\exp(2i\Phi(I)) \cdot {\cal A}_{{\cal N}(I)}(x,t),
$$
defined by
\begin{equation}
\label{eq:Akh_CS}
{\cal A}_{{\cal N}(I)}(x,t)   =\exp(2it) \frac{\hat\theta_{2{\cal N}(I)}(\hat Z_{+}(x,t)|B)}{\hat\theta_{2{\cal N}(I)}(\hat Z_{-}(x,t)|B )}, 
\end{equation}
where (we keep only the ``p-visible'' components in the sum):
\begin{equation}
\label{eq:theta-hyper3}
\hat\theta_{2{\cal N}(I)}(\hat Z|B) = \!\!\!\!\!\!\!\!\!\!\!\!\!\!\!\!\!\!\!\!\!\!  \sum\limits_{\begin{array}{c}\hat n_j\in\{-1,1\}\\ j=1,\ldots,2{\cal N}(I)\end{array}}\!\!\!\!\!\!\!\!\!\!\!\!\!\!
\exp{\left[\!\! \sum\limits_{\begin{array}{c}l,m=1 \\ l\ne m\end{array}}^{2{\cal N}(I)}\!\!\!\!\!\! 
\log \left|\frac{\sin\left(\frac{\hat\phi_{s_l(I)}-\hat\phi_{s_m(I)}}{2} \right)}{\cos\left(\frac{\hat\phi_{s_l(I)}+\hat\phi_{s_m(I)}}{2} \right) }\right| \frac{\hat n_l\hat n_m}{4} +  \pi i \sum\limits_{l=1}^{2{\cal N}(I)}  \hat n_l \hat Z_l\right]},
\end{equation}
$$
\left(\hat Z_{-}(x,t)\right)_j = \left(\hat z_{-}(x,t)\right)_{s_j(I)}, \ \ j=1,\ldots, 2{\cal N}(I),$$ 
,
$$
(\hat Z_{+}(x,t))_k = (\hat Z_{-}(x,t))_k - \frac{1}{2} -\frac{\hat\phi_{s_k(I)}}{\pi}, 
$$
$$
\hat z_{-}(x,t) = \vec z_{-} (x,t) - \sum\limits_{k=1}^{2N} \left[\lfloor w_k(t) \rfloor +\frac{st_k(I)}{2} \right] \vec b_k ,
$$
\beq\label{phases}
\Phi(I) = \sum\limits_{k=1}^{N} \left[ 2\lfloor w_k(t) \rfloor + st_k(I) \right] \hat\phi_k , 
\eeq
and $\vec b_k$ is the $k^{th}$ column of matrix $B$.
\end{theorem}
Therefore, in each time interval $I$ of the partition, we approximate $u(x,t)$, up to $O(\epsilon^p)$, $0< p \le 1$ corrections, by elementary functions explicitly defined in terms of Cauchy data. For instance, in Figure~\ref{fig:3modes} we use the above analytic formulas to construct the leading order solution, in the case of 3 unstable modes ($L=10$), and $p=1/2$, in very good agrement with the corresponding numerical experiment.

\begin{remark}
To the best of our knowledge, the exact ${\cal N}$-soliton solution of Akhmediev type has the following history. The representation (\ref{eq:Akh_CS}) was obtained by Its, Rybin and Sall in \cite{ItsRybinSall}, as a degeneration of the finite-gap formulas. The determinant form of this solution and a discussion on the connection between these two representations are also provided in \cite{ItsRybinSall}. Let us remark that formulas (\ref{eq:Akh_CS}) can be obtained from the Hirota ${\cal N}$-soliton solution \cite{Hirota} of defocusing NLS by a complex rotation \cite{CaliniShober1}. Determinant formulas for the ${\cal N}$-soliton solution over the zero background for the self-focusing case were provided in the book of Faddeev and Takhtadjan \cite{FT}, and the determinant formula for the ${\cal N}$-soliton solution over an arbitrary background, including the so-called super-regular solitons, was constructed by Zakharov and Gelash \cite{ZakharovGelash1}; see also \cite{ZakharovGelash2}, \cite{ZakharovGelash3}.
\end{remark}

\begin{remark}
Formula (\ref{eq:theta-hyper3}) involves summations over $4^{\cal N}, \ {\cal N}\le N$ terms of $O(1)$, growing exponentially with $\cal N$ (f.i., if $N=5$, the sum involves about $10^3$ terms). It follows that these sums could generate contributions comparable with $O(|\log\eps|)$, for reasonable values of $\eps$ coming from physics or from numerical simulations, and the results would become less reliable. Therefore the number $N$ of unstable modes must be sufficiently small to avoid this problem.   
\end{remark}

\begin{remark}
In this paper we use a more symmetric notation with respect to the one used in our previous works on this subject \cite{GS1}-\cite{GS5} (see, f.i., equations (\ref{def_alpha_beta}). We also use a different normalization for the theta-functions: in the present text it coincides with the one used in \cite{Fay}, \cite{BBEIM}, \cite{Dubrovin2}, while, in the previous works, we used the normalization from the book \cite{Mum1983}. 
\end{remark}

\section{Periodic problem for the Nonlinear Schr\"odinger Equation}
\label{sec:periodicNLS}

Let us recall the basic facts about the periodic theory of the Nonlinear Schr\"odinger (NLS) equation.

The NLS equation has two real forms:
\begin{equation}
\label{eq:rNLS}
\begin{array}{ll}
i u_t +u_{xx} - 2 |u|^2 u=0, \ \ u=u(x,t)\in\CC, & \mbox{defocusing NLS} \\ 
i u_t +u_{xx} + 2 |u|^2 u=0, \ \ u=u(x,t)\in\CC, & \mbox{self-focusing NLS}.
\end{array}
\end{equation}
In nonlinear optics, both models describe media with refractive index depending on the electric filed. If the refractive index decreases (increases) in the presence of the electromagnetic wave, we have defocusing (respectively self-focusing) NLS equation. Both forms can be treated as real reductions of the complex NLS equation 
\beq\label{eq:cNLS}
\left\{
\begin{array}{l}
i q_t +q_{xx}+2 q^2 r=0, \\
-i r_t +r_{xx}+2 q r^2=0, \ \ q=q(x,t), \ \ r=r(x,t) \ \  \in\CC,
\end{array}
\right.
\eeq
where
\begin{align}
&q(x,t) = u(x,t), \ \ r(x,t) = -\overline{u(x,t)} & &\mbox{defocusing case}\\
&q(x,t) = u(x,t), \ \ r(x,t) = \overline{u(x,t)} & &\mbox{self-defocusing case}
\end{align}
The integration of the NLS equation is based on the zero-curvature representation found by Zakharov and Shabat in \cite{ZakharovShabat}.   
A pair of functions $q(x,t)$, $r(x,t)$ satisfies the complex NLS equation (\ref{eq:cNLS}) if and only if the following pair of linear problems is compatible: 
\begin{align}
\label{eq:zc_x}
\vec\Psi_x(\lambda,x,t)&=U(\lambda,x,t)\vec\Psi(\lambda,x,t),\\
\label{eq:zc_t}
\vec\Psi_t(\lambda,x,t)&=V(\lambda,x,t)\vec\Psi(\lambda,x,t),
\end{align}
$$
U=\left [\begin {array}{cc} -i \lambda & i q(x,t)
\\\noalign{\medskip} i r(x,t) & i \lambda\end {array}
\right ],
$$
where
$$
V(\lambda,x,t)= \left[\begin {array}{cc} -2 i \lambda^2 + i q(x,t) r(x,t) & 2 i \lambda q(x,t) - q_x(x,t)
\\\noalign{\medskip} 2 i \lambda r(x,t) + r_x(x,t) & 2 i \lambda^2- i q(x,t) r(x,t)  
\end {array}
\right ],
$$
and
$$
\vec\Psi(\lambda,x,t)= \left [\begin {array}{c} \Psi_1(\lambda,x,t) \\
\Psi_2(\lambda,x,t) \end {array}\right ].
$$
The first equation of the zero-curvature representation (\ref{eq:zc_x}) can be rewritten as the following spectral problem
\begin{equation}
\label{eq:lp-x2}
\LG\vec\Psi(\lambda,x,t)=\lambda \vec\Psi(\lambda,x,t),
\end{equation}
where 
$$
\LG=\left[ \begin {array}{cc} i\partial_x & q(x,t) \\ -  r(x,t) & -i\partial_x \end {array}\right].
$$
The main tool for constructing periodic and quasiperiodic solutions of soliton systems is the finite-gap method, invented by S.P. Novikov \cite{Novikov} for the periodic Korteweg - de Vries (KdV) problem. Finite-gap solutions of the NLS equation were first constructed by Its and Kotljarov \cite{ItsKotlj}.

Let us recall the principal facts about the periodic direct and inverse spectral transform for the 1-dimensional Dirac operator (\ref{eq:lp-x2}).

Consider a fixed time $t=t_0$. Let $q(x)$, $r(x)$ denote the Cauchy data $q(x)=q(x,t_0)$, $r(x)=r(x,t_0)$, $q(x+L)=q(x)$, $r(x+L)=r(x)$. The direct spectral transform associates to these Cauchy data the following spectral data: 
\begin{enumerate}
\item The spectral curve $\Gamma$, i.e., the Riemann surface for the Bloch eigenfunctions.
\item The divisor, i.e., the set of eigenvalues for an auxiliary Dirichlet-type spectral problem. 
\end{enumerate}

\textbf{1. The spectral curve.} The Bloch functions of $\LG$ are defined as the common eigenfunctions of  $\LG$ and of the translation operator (the periodicity of  $\LG$ means that it commutes with the translation by the basic period $L$; therefore $\LG$ and the translation operator have sufficiently many common eigenfunctions):
\begin{eqnarray}
\label{eq:bloch1}
\LG  \vec\Psi(x,t_0) =\lambda \vec\Psi(x,t_0),\hphantom{aaa\gamma \in \Gamma}\nonumber\\
\vec\Psi(x+L,t_0)=\kappa \vec\Psi(x,t_0), \ \ \gamma \in \Gamma .
\end{eqnarray}
Equivalently, the Bloch functions are eigenfunctions of the monodromy matrix  $\hat T(\lambda,x_0,t_0)$ defined by:
$$
\hat T(\lambda,x_0,t_0)=\hat\Psi(\lambda,x_0+L,t_0),
$$
where  the $2\times2$ invertible matrix $\hat\Psi(\lambda,x,t_0)$ denotes the solution of the matrix equation 
$$
\LG\hat\Psi(\lambda,x,t_0)=\lambda\hat\Psi(\lambda,x,t_0),
$$
with the following initial condition
$$
\hat\Psi(\lambda,x_0,t_0)=\left [\begin {array}{cc} 1 & 0
\\ 0 & 1 \end {array}\right].
$$
The monodromy matrix $\hat T(\lambda,x_0,t_0)$ is holomorphic in $\lambda$ in the whole complex plane. It follows that the eigenvalues and eigenvectors of $\hat T(\lambda,x_0,t_0)$ (i.e., the Bloch multiplier $\kappa$ and the Bloch eigenfunction $\vec\Psi(x,t_0)$) are defined on a two-sheeted covering $\Gamma(x_0,t_0)$ of the $\lambda$-plane; therefore we shall write $\kappa(\gamma)$,  $\vec\Psi(\gamma,x,t_0)$, $\gamma\in\Gamma$. 

The Riemann surface $\Gamma(x_0,t_0)$ is called the \textbf{spectral curve}. If the potentials $q(x,t)$, $r(x,t)$ satisfy the NLS equation (\ref{eq:cNLS}), then the monodromy matrices corresponding to different $x_0$ and $t_0$ coincide up to conjugation; therefore $\Gamma(x_0,t_0)$ does not depend on $x_0$ and $t_0$, and will be denoted by $\Gamma$ in the remaining part of the text. In particular, this means that the curve $\Gamma$ provides an infinite set of conservation laws for the NLS hierarchy. This set is complete, and the standard local conservation laws can be easily obtained as expansion coefficients of $\Gamma$ near infinity. 

The spectrum  of the problem (\ref{eq:lp-x2}) in $L^2(\RR)$ is always continuous, and it is defined by the following property: $\lambda\in\CC$ belongs to the spectrum of $\LG$ if and only if equation (\ref{eq:lp-x2}) admits a solution growing not faster then some polynomial for both $x\rightarrow -\infty$ and $x\rightarrow +\infty$. The ends of the spectral intervals coincide with the branch points of $\Gamma$. In the paper \cite{Novikov} the spectral curve was introduced for the real Schr\"odinger operator, where the relation between the spectral curve and the classical spectrum is especially simple: the spectrum in $L^2(\RR)$ is the set of intervals in the real line bounded by the branch points of  $\Gamma$.

We use the following notation: if $\gamma\in\Gamma$ is a point of our spectral curve, then $\lambda(\gamma)$ denotes the projection of the point $\gamma$ to the $\lambda$-plane. 

The multivalued function 
\begin{equation}
\label{eq:bloch2}
p(\gamma)=\frac{1}{iL}\log(\kappa(\gamma))
\end{equation}
is called \textbf{quasimomentum.} Its differential $dp(\gamma)$ is well-defined and meromorphic on $\Gamma$ with two simple poles at the infinity points, and all periods of $dp$ are real. The trace of matrix $U(\lambda,x,t)$ is equal to zero; therefore 
$$
\det \hat T(\lambda,x_0,t_0) \equiv 1, 
$$
and if $\lambda(\gamma_1)=\lambda(\gamma_2)$, then $\kappa(\gamma_1)\kappa(\gamma_2)=1$.

The ``classical'' spectrum of $\LG$ in $L^2(\RR)$ coincides with the projection of the set $\{\gamma\in\Gamma, |\kappa(\gamma)|=1\}$ to the $\lambda$-plane, or, equivalently, is defined by the condition:
$$
\Im p(\gamma)=0.
$$

The branch points of $\Gamma$ coincide with the ends of the spectral zones.

\begin{definition}
A pair of potentials $q(x)$, $r(x)$ is called \textbf{finite-gap} if the spectral curve $\Gamma$ is algebraic, i.e., it has only a finite number of branch points and non-removable double points. In the real case $r(x)=\pm\overline{q(x)}$, the second requirement is fulfilled automatically, and it is sufficient to demand that the number of branch points be finite. 
\end{definition}

\begin{remark}
The analytic properties of the Bloch eigenfunction for the 1-dimensional Schr\"odinger operator in the domain of complex energies were first studied in the paper by  Kohn \cite{Kohn}.
\end{remark}

\textbf{2. Divisor}  The auxiliary spectrum is defined as the set of points $\gamma\in\Gamma$ such 
that the first component of the Bloch eigenfunction is equal to 0 at the point $x_0$.
\begin{eqnarray}
\label{eq:dir1}
\LG\vec\Psi(\gamma,x,t)= \lambda(\gamma) \vec\Psi(\gamma,x,t),\nonumber\\
\vec\Psi(\gamma,x+L,t))=\kappa(\gamma) \vec\Psi(\gamma,x,t),\\
\Psi_1(\gamma,x_0,t_0)=0.\nonumber
\end{eqnarray}
Equivalently, the auxiliary spectrum coincides with the set of zeroes of the first component of the Bloch 
eigenfunction:
\begin{equation}
\label{eq:dir2}
\Psi_1(\gamma,x,t)=0;
\end{equation}
therefore it is called the \textbf{divisor of zeroes.} The zeroes of $\Psi_1(\gamma,x,t)$ depend on $x$ and $t$. The $x$, 
$t$ dynamics becomes linear after the Abel transform (for KdV this fact was first established by Dubrovin \cite{Dubrovin}, Its and Matveev \cite{ItsMatveev}). 

\begin{remark}
Let us point out that, in some papers dedicated to the finite-gap NLS solutions, 
a different auxiliary problem is used. Namely, one imposes the following symmetric boundary condition:
\begin{equation}
\label{eq:dir3}
\Psi_1(\lambda,x_0,t_0)+\Psi_2(\lambda,x_0,t_0) = \Psi_1(\lambda,x_0+L,t_0)+\Psi_2(\lambda,x_0+L,t_0)=0.
\end{equation}
This approach has the following advantage: all divisor points are located in a compact area of the spectral curve,  
but it requires one extra divisor point and increases the complexity of the formulas.
\end{remark}
 
The wave propagation in focusing and defocusing media is substantially different from the physical point of view. This difference means that the analytic properties of the solutions are also substantially different (this topic is well-presented in the paper by Previato \cite{Prev}).
\begin{enumerate}
\item In the defocusing case:
\begin{itemize}
\item Operator $\LG$ is self-adjoint, and its spectrum is real; 
\item All branch points of $\Gamma$ are real and simple;
\item If the normalization (\ref{eq:dir3}) is used, each non-empty spectral gap contains exactly one divisor point;
\item The defocusing NLS equation possesses regular and singular finite-gap solutions. 
\end{itemize}
From the analytic point of view, the theory of defocusing NLS is analogous to the theory of the real Korteweg-de Vries equation.
\item In the self-focusing case:
\begin{itemize}
\item Operator $\LG$ is non-self-adjoint, and the matrix $U(\lambda,x,t)$ is skew-hermitian for real $\lambda$. Therefore, for each real $\lambda$, the monodromy matrix is unitary. It means that the whole real line lies in the $L^2(\RR)$ spectrum of $\LG$; in addition, generically, the $L^2(\RR)$ spectrum of $\LG$ contains some arcs in the complex domain.
\item All branch points of $\Gamma$ are complex, and a finite number of them may be of high odd multiplicity. All real double points of $\Gamma$ are removable (the monodromy matrix has two different eigenfunctions), but complex double points may be non-removable, and such points are associated with homoclinic orbits. For sufficiently regular data, the number of complex double points is always finite. The curve $\Gamma$ is real; i.e., it is invariant with respect to the complex conjugation. 
\item The characterization of the divisor is less explicit, and it can be provided in terms of Cherednik differentials \cite{Cher}. 
\item All finite-gap solutions are automatically regular \cite{Cher}.
\end{itemize}
In contrast with the defocusing case, the self-focusing case is much richer and much more interesting from the analytic point of view.
\end{enumerate}

For generic smooth NLS Cauchy data $q(x,t_0)$, $r(x,t_0)$, the surface $\Gamma$ has infinite genus, but for large values of the spectral parameter, the branch points of $\Gamma$ form close pairs (see \cite{DjMit} for analytic estimates extending the results of \cite{Hochstadt} to the 1-d Dirac operator), and one can construct an arbitrarily good finite-gap approximation for a given smooth potentials by merging all close pairs of branch points except a finite number. ``Naive'' merging does not respect the spatial periodicity and provides a local in $x$ approximation only. In the defocusing case, a period preserving approximation may be obtained using a minor modification of the Marchenko-Ostrovskii approach \cite{MO}. In the focusing case, the analytic properties of the quasimomentum are more complicated, and, to construct a pure periodic finite-gap approximation, one can use either the isoperiodic deformation (Grinevich-Schmidt \cite{GrSch}), or a proper adaptation of the Krichever's technique from the paper \cite{Krichever2}.

Therefore, in the periodic problem for soliton equations, the role of finite-gap potentials can be compared with that of finite Fourier series in the theory of linear PDEs. 

The solution of the inverse problem in the finite-gap case (the genus of $\Gamma$ is equal to $g<\infty$) is provided by the theta-functional formula (\cite{ItsKotlj}, see also \cite{Prev}):
 
\begin{equation}
\label{eq:pot1_1}
u(x,t)={\cal C} \ \exp({\cal U}x + {\cal V} t) \ \frac{\theta( \vec A(\infty_{-}) -\vec U_1 x - \vec U_2 t -\vec A(\DD) -\vec K |B)}{\theta(\vec A(\infty_+) -\vec U_1 x - \vec U_2 t -\vec A(\DD) -\vec K |B)},
\end{equation}
where $\cal C$, $\cal U$, $\cal V$ are constants defined in terms of the spectral curve,  $\vec A(\DD)$, $\vec A(\infty_{+})$,  $\vec A(\infty_{-})$ are the Abel transforms of the divisor, and of the infinity points of $\Gamma$ respectively, $\vec K$ is the so-called vector of Riemann constants, $B$ is the Riemann period matrix for $\Gamma$, $\vec U_1$, $\vec U_2$ are the vectors of the $b$-periods for the quasimomentun and quasienergy differentials, respectively, (see formulas (\ref{eq:U1}),  (\ref{eq:U2})), and $\theta(z|B)$ denotes the Riemann theta-function of genus $g$ 
\begin{equation}
\label{eq:theta0}
\theta(z|B)=\sum\limits_{\begin{array}{cc}n_j\in\ZZ \\ j=1,\ldots,g\end{array}}
\exp{\left[ \frac{1}{2} \sum_{j,k=1}^g b_{jk}n_j n_k +  \sum_{j=1}^g n_j z_j  \right] }.
\end{equation}
where $b_{jk}$ are the components of matrix $B$. More informations about theta-functional formulas can be found in \cite{Dubrovin2}, \cite{BBEIM}. Explicit approximate formulas for the parameters appearing in (\ref{eq:pot1_1}),(\ref{eq:theta0}), in the special Cauchy problem of anomalous waves, are provided in Section~\ref{sec:sec4}.

The normalization (\ref{eq:theta0}) implies the following periodicity properties:
\begin{align}
\theta\left(\vec z+ \vec a_l |B \right)&=\theta\left(\vec z |B \right) \nonumber\\
\label{eq:theta-per} 
\theta\left(\vec z+ \vec b_l |B \right)&=\theta\left(\vec z |B \right) \exp\left(-\frac{1}{2} b_{ll}- z_l \right),
\end{align}
where $l=1,\ldots,2N$.

\section{Spectral transform for the Cauchy problem of  the anomalous waves}
\label{sec:sec4}

If one uses the focusing NLS equation as a mathematical model for anomalous waves, special initial data (\ref{Cauchy})-(\ref{Fourier})  are considered. We show that the presence of a small parameter $\epsilon$ in this problem allows one to construct a good approximation for the solutions in terms of elementary functions.

To construct the direct spectral transform for this problem, it is convenient to write
$$
\LG=\LG_0+\eps \LG_1, \ \ \LG_0 = \left[ \begin {array}{cc} i\partial_x & 1 \\ -1  & -i\partial_x \end {array}\right], \ \ 
\LG_1 = \left[ \begin {array}{cc} 0 & v(x) \\ -\overline{v(x)}  & 0 \end {array}\right],
$$
and calculate the spectral data for $\LG$ using the standard perturbation theory near the spectral data for $\LG_0$. The leading order formulas for the spectral curve and the divisor were provided in the recent paper \cite{GS1} of the authors. Let us present a brief review of these results.

\subsection{The spectral data for the unperturbed operator}
\label{sec:sec4.1}

The unperturbed spectral curve $\Gamma_0$  for $\LG_0$ is rational, and a point $\gamma\in\Gamma_0$ is a pair of complex numbers
$\gamma=(\lambda,\mu)$ satisfying the following quadratic equation:
$$
\mu^2=\lambda^2+1.
$$
The Bloch eigenfunctions for the operator $\LG_0$ can be easily calculated explicitly:
\begin{equation}
\label{eq:bloch3}
\psi^{\pm}(\gamma,x)=\left[\begin {array}{c} 1 \\ \lambda(\gamma)\pm \mu(\gamma) \end {array} \right ] e^{\pm i\mu(\gamma) x},
\end{equation}
$$
\LG_0 \psi^{\pm}(\gamma,x) = \lambda(\gamma)  \psi^{\pm}(\gamma,x).
$$
These eigenfunctions are periodic (antiperiodic) if and only if $\frac{L}{2\pi}\mu\in\ZZ$ is an even (an odd) integer. Let us introduce the following notation:
$$
\mu_n = \frac{\pi n}{L}, \ \ \lambda_n= \sqrt{\mu_n^2-1}, \ \ \Re{\lambda_n}+\Im{\lambda_n}>0, \ \  
n=0,1,2,\ldots\infty.
$$
Then the periodic and antiperiodic spectral points of $\LG_0$ are (see Figure 5):
$$
\{\pm\lambda_n\}, \ \ n=0,1,2,\ldots\infty.
$$
By analogy with \cite{Krichever2}, \cite{Krichever3}, these points are called \textbf{resonant point}, because, under generic small perturbations, they split into pairs of branch points.
\begin{figure}[H]
\centering
\mbox{\epsfxsize=6cm\epsffile{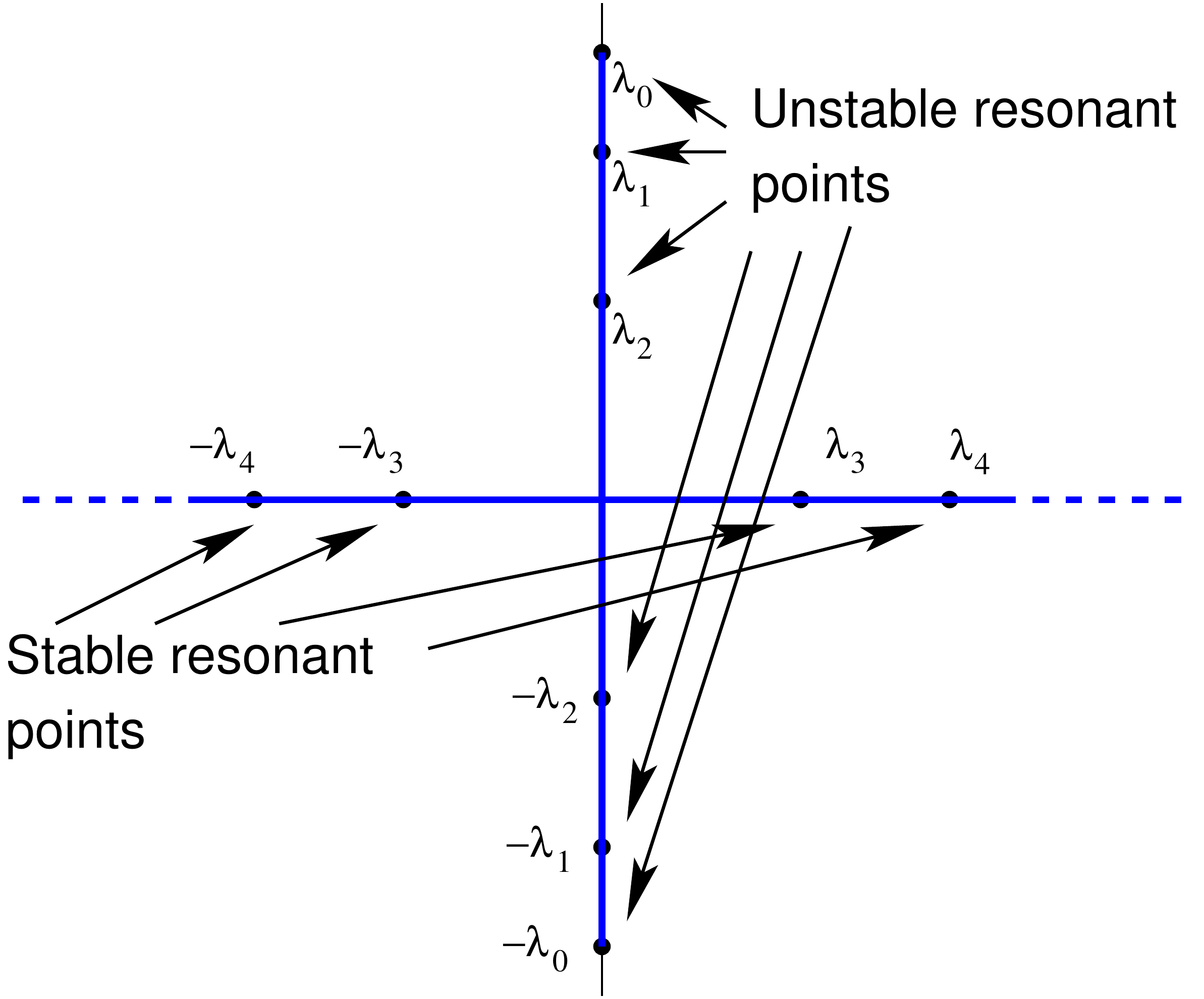}}
\caption{The spectrum of the unperturbed operator $\LG_0$.}
\end{figure}
We also use the following notation:
$$
\lambda_{-n} = \lambda_n, \ \ \mu_{-n} = -\mu_n, \ \ n\ge 0.
$$
Let us recall that we assume $L\ne\pi n$, where $n\in\NN_+$; therefore the point $\lambda=0$ is not resonant. The case of resonant $\lambda=0$ requires a special treatment, and we plan to provide it in the future.

We have the following basis of eigenfunctions for the periodic and antiperiodic problems:
\begin{equation}
\label{eq:basis1}
\psi^{\pm}_{n}=\left[\begin {array}{c} 1 \\ \mu_n\pm\lambda_n \end {array} \right ] e^{i{\mu_n}x}, \ \ n\in\ZZ,
\end{equation}
$$
\LG_0\psi^{\pm}_{n}=\pm\lambda_{n} \psi^{\pm}_{n}.
$$
The curve $\Gamma_0$ has two branch points $E_0=i$, $\overline{E_0}=-i$ corresponding to $n=0$. If $n>0$, there is no 
branching at the resonant points $\pm\lambda_n$, but the monodromy matrix becomes diagonal with coinciding eigenvalues:
$$
\hat T(\pm\lambda_n,0,0)= \left [\begin {array}{cc} (-1)^n &  0
\\ \noalign{\medskip} 0 & (-1)^n \end {array}
\right ].
$$
Resonant points are also the eigenvalues of the Dirichlet problem (\ref{eq:dir1}), i.e., \textbf{the divisor points for the unperturbed problem}, 
and the Dirichlet eigenfunctions are given by
$$
\psi^{\mbox{Dir}}_{n}(x)=\psi^{+}_{n}(x)-\psi^{+}_{-n}(x), \ \  \psi^{\mbox{Dir}}_{-n}(x)=\psi^{-}_{n}(x)-\psi^{-}_{-n}(x),  \ \ n>0.
$$

Taking into account that the squared eigenfunctions provide the proper basis for the linearized theory, we notice that the  
Fourier modes correspond the the points $\mu_n=k_n/2$. These modes are unstable if the corresponding $\lambda_n$ are 
imaginary ($|\mu_n|<1$)  and stable if the  $\lambda_n$ are real ($|\mu_n|\ge 1$); see the Introduction. Therefore the spectral points 
with  $|n|<\frac{L}{\pi}$ are \textbf{unstable}, and spectral points with $|n|\ge\frac{L}{\pi}$  are \textbf{stable}. For the unstable modes we have:
\begin{equation}
\label{eq:lmphi}
\lambda_j = i\cos(\phi_j), \ \ \mu_j = \sin(\phi_j), \ \ j=1,\ldots,N,
\end{equation}
where the angles $\phi_j$  are the same as in (\ref{def_angle0}).

\subsection{The spectral data for the perturbed operator}
\label{sec:sec4.2}

To calculate the perturbed spectral curve, we develop the perturbation theory for the periodic and antiperiodic problems 
using the basis (\ref{eq:basis1}).

It is also convenient to introduce the following notation where $n\ge 1$:
\begin{align}
\label{def_alpha_beta_2}
\alpha_n =(\mu_n-\lambda_n)  \overline{c_n}-(\mu_n+\lambda_n) c_{-n}, \ \ \beta_n =(\mu_n+\lambda_n) \overline{c_{-n}}-(\mu_n-\lambda_n) c_{n},\\
\tilde{\alpha}_n =(\mu_n+\lambda_n) \overline{c_n}-(\mu_n-\lambda_n) c_{-n}, \ \ \tilde{\beta}_n =(\mu_n-\lambda_n)\overline{c_{-n}}-(\mu_n+\lambda_n) c_{n},
\end{align}
which is consistent with (\ref{def_alpha_beta}) since, for the unstable modes, $e^{\pm i\phi_n}=\mu_n\pm\lambda_n$.

Let us introduce the following notation: $|l_{\pm}>$ denotes the basic vector $\psi^{\pm}_l$, $<l_{\pm}|$ denote the adjoint 
basis:
$$
<l_{+}|m_{+}> = \delta_{lm}, \ \ <l_{-}|m_{-}> = \delta_{lm}, \ \ <l_{+}|m_{-}> = <l_{-}|m_{+}>=0.
$$
For an arbitrary periodic perturbation, the matrix elements of $\LG_1$ can be written in the following form: 
$$
<m_{+}|\LG_1|l_{+}> = \frac{c_{(m-l)/2}(\lambda_{m}-\mu_{m})(\lambda_l+\mu_l)-\overline{c_{-(m-l)/2}}}{2\lambda_{m}},
$$
$$
<m_{-}|\LG_1|l_{+}> = \frac{c_{(m-l)/2}(\lambda_{m}+\mu_{m})(\lambda_l+\mu_l)+\overline{c_{-(m-l)/2}}}{2\lambda_{m}},
$$
$$
<m_{+}|\LG_1|l_{-}> = \frac{c_{(m-l)/2}(\lambda_{m}-\mu_{m})(-\lambda_l+\mu_l)-\overline{c_{-(m-l)/2}}}{2\lambda_{m}},
$$
$$
<m_{-}|\LG_1|l_{-}> = \frac{c_{(m-l)/2}(\lambda_{m}+\mu_{m})(-\lambda_l+\mu_l)+\overline{c_{-(m-l)/2}}}{2\lambda_{m}}.
$$
Here we use a slightly non-standard notation: $<f|\LG|g>$ denotes the matrix element for both orthogonal and 
non-orthogonal bases. We also assume that $c_{-(m-l)/2}=0$, if $m-l$ is odd.

Using the standard perturbation theory (see \cite{GS1} for details), one shows that the resonant points $\lambda_n$ and $-\lambda_{n}$ generically split into the pairs of  branch points $\{ E_{2n-1},E_{2n}\}$ and $\{ \tilde{E}_{2n-1},\tilde{E}_{2n}\}$, where:
\begin{align}
\label{eq:bps2}
E_0 &= i + O(\epsilon^2) \nonumber \\
E_{l}& =\lambda_n\mp\frac{\epsilon}{2\lambda_n}\sqrt{\alpha_n\beta_n}+O(\epsilon^2),  \ \ l=2n-1,2n,\\
\tilde{E_{l}} &=-\lambda_n\pm\frac{\epsilon}{2\lambda_n}\sqrt{\tilde\alpha_n\tilde\beta_n}+O(\epsilon^2),  \ \ l=2n-1,2n \hphantom{l=2n-1,2n}. \nonumber
\end{align}
(see Figure 6). Here we assume that $\Re\sqrt{\alpha_n\beta_n}\ge0$, $\Re\sqrt{\tilde\alpha_n\tilde\beta_n}\ge0$ for the unstable points, and  $\Im\sqrt{\alpha_n\beta_n}<0$, $\Im\sqrt{\tilde\alpha_n\tilde\beta_n}<0$ for the stable points. To estimate the variation of $\lambda_0$, we also used the constraint $c_0=0$.

For the perturbations of the unstable points we have:
\begin{equation}
\label{eq:bps3}
\tilde{E_{l}}=\overline{E_{l}}.
\end{equation}

\begin{figure}[H]
\centering
\mbox{\epsfxsize=10cm\epsffile{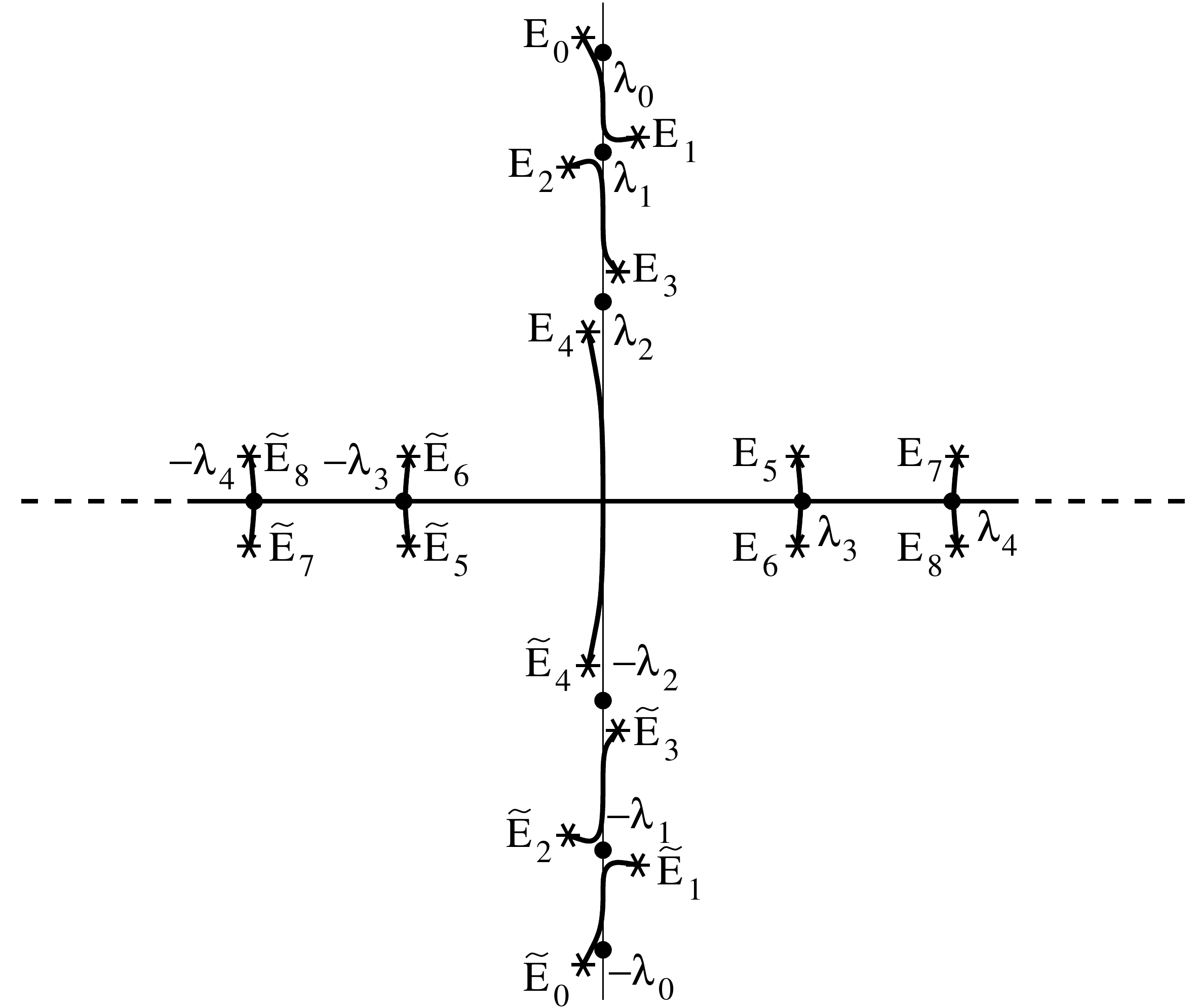}}
\caption{The spectrum of the perturbed operator $\LG=\LG_0+\eps \LG_1$.}
\end{figure}

Let us define the following enumeration for the  unperturbed divisor points 
$\gamma_n=(\lambda^{\mbox{\small div}}_n, \mu^{\mbox{\small div}}_n)$,  $n\ne 0$:
$$
\lambda^{\mbox{\small div}}_n=\left\{\begin{array}{ll} \lambda_n, & n>0, \\  -\lambda_n, & n<0, \end{array}\right. \ \ 
\mu^{\mbox{\small div}}_n=\mu_n.
$$
The calculation of the divisor positions up to $O(\epsilon^2)$ corrections uses the same perturbation theory with the following modification. In contrast with the branch points, the Bloch multipliers for the Dirichlet spectrum are generically different from $\pm 1$; moreover their absolute values do not have to be equal to 1. Therefore we use the additional constraint that the first component of the Dirichlet eigenfunction vanishes at the boundary of the interval to determine simultaneously the variation of $\lambda$ and the variation of the  Bloch multiplier $\kappa$, and we obtain (see \cite{GS1}):

\begin{equation}
\label{eq:div_alpha1}
\begin{array}{ll}
\lambda(\gamma_n) = \lambda_n+\frac{\epsilon}{4\lambda_n}\left[(\mu_n+\lambda_n) \alpha_n+ (\mu_n-\lambda_n) \beta_n\right]+O(\epsilon^2), \\
p(\gamma_{n})= \frac{\epsilon}{4\mu_n}\left[(\mu_n+\lambda_n) \alpha_n-(\mu_n-\lambda_n)\beta_n\right]+O(\epsilon^2),
\end{array}
\end{equation}
and
\begin{equation}
\label{eq:div_alpha2}
\begin{array}{ll}
\lambda(\gamma_{-n}) =-\lambda_n-\frac{\epsilon}{4\lambda_n}\left[(\mu_n-\lambda_n) \tilde\alpha_n+ (\mu_n+\lambda_n)\tilde\beta_n\right]+O(\epsilon^2), \\
p(\gamma_{-n})= \frac{\epsilon}{4\mu_n}\left[ (\mu_n-\lambda_n)\tilde\alpha_n- (\mu_n+\lambda_n)\tilde\beta_n\right]+O(\epsilon^2).
\end{array}
\end{equation}

\section{The ``unstable part'' of the spectral curve}
\label{sec:sec5}

A generic small perturbation of the constant solution generates an infinite genus spectral curve. Of course, it is
natural to approximate it by a finite-genus curve. The perturbations corresponding to stable resonant points remain 
of order $\epsilon$ for all $t$ and can be well-described by the linear perturbation theory. Therefore one can close 
all gaps associated with stable points, obtaining a finite-gap approximation of the spectral curve. 

\begin{remark}
This finite-gap approximation is rather non-standard. Usually, one keeps the gaps associated with sufficiently large harmonics and close the gaps corresponding to the small ones; therefore the genus of the approximating curve depends on the perturbation. In our problem, all harmonics of the perturbation are small, and the genus of the approximating curve is determined by the number of unstable modes and does not depend on the Cauchy data.

\end{remark}

\subsection{Finite genus approximating curve}

Starting from this point, we shall use the following $2N$-gap approximation of the spectral curve: we open only the resonant points associated with the unstable modes, and do not perturb the stable double point. 

Our next step is to calculate the leading order approximation of the algebro-geometrical data. 

\begin{figure}[H]
\centering
\includegraphics[width=0.5\textwidth]{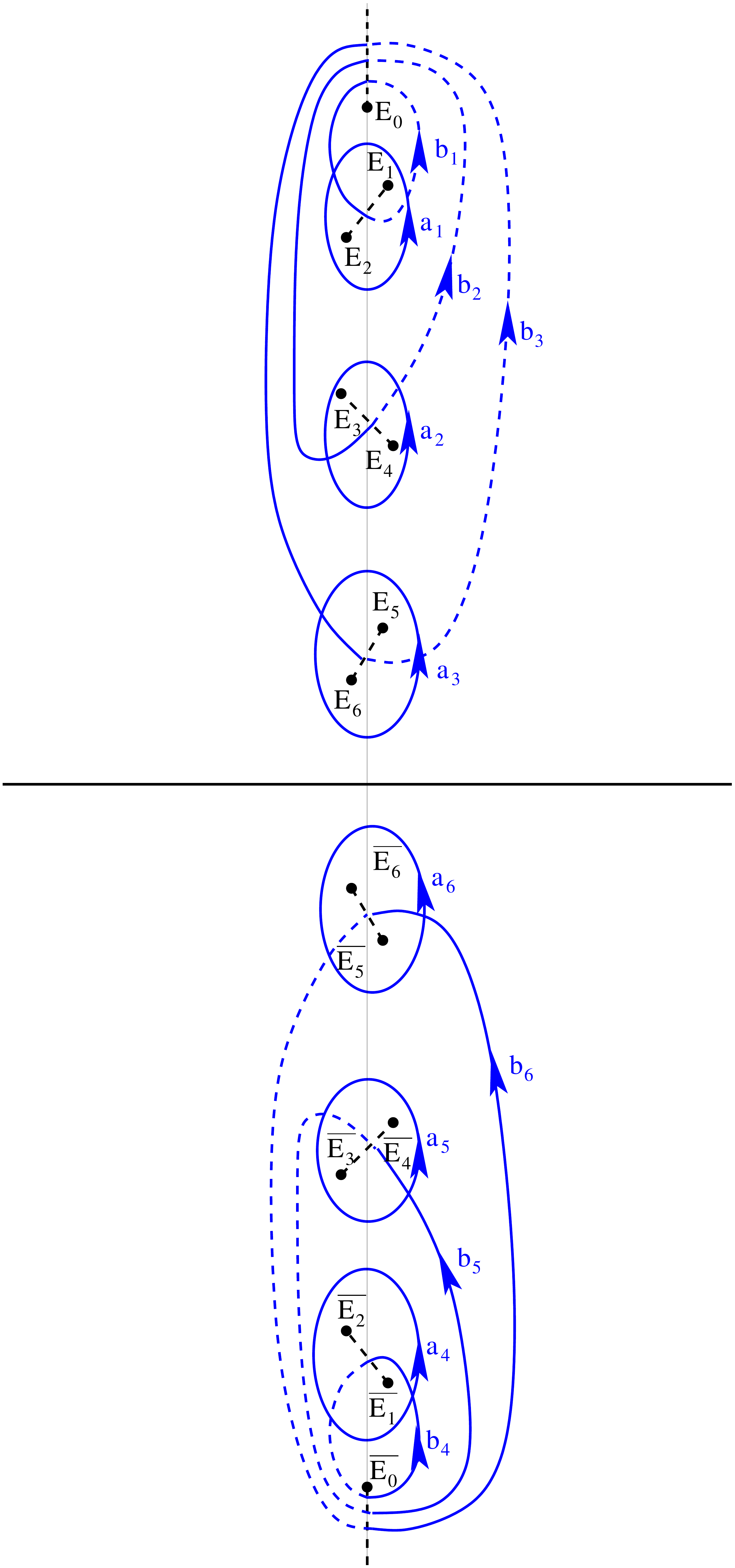}
\caption{\label{fig:basis1}\small{\sl The system of cuts and basis of cycles used in our paper for $N=3$. The objects on Sheet~1 are drawn in solid lines, the objects on Sheet~2 are drawn in dashed lines. All cycles $a_j$ lie on Sheet~1.}}
\end{figure}

Let us introduce the following notations:
$$
\hat\lambda_j =\left\{\begin{array}{ll} \lambda_j, & \mbox{if} \ \ 1\le j\le N, \\ & \\
\overline{\lambda_{j-N}}, &  \mbox{if} \ \ N+1\le j\le 2N, \end{array} \right.
$$
$$
\hat\mu_j =\left\{\begin{array}{ll} \mu_j, & \mbox{if} \ \ 1\le j\le N, \\ & \\
\mu_{j-N}, &  \mbox{if} \ \ N+1\le j\le 2N. \end{array} \right.
$$
$$
\hat\gamma_j=\left\{\begin{array}{ll} \gamma_j & \mbox{if} \ \ 1\le j\le N \\ & \\
\gamma_{-j} &  \mbox{if} \ \ N+1\le j\le 2N \end{array} \right.
$$
We have $4N+2$ branch points:
$$
E_j, \ \ j=0,\ldots, 2N, \ \ \overline{E_j}, \ \ j=0,\ldots, 2N,
$$
and the curve $\Gamma$ is defined by:
\begin{equation}
\label{eq:curve1}
\mu^2=\prod\limits_{j=0}^{2N}(\lambda-E_j)(\lambda-\overline{E_j}).
\end{equation}
We also define
$$
\mu^{(0)}(\lambda)=\sqrt{\lambda^2+1},
$$
$$
z_n=\frac{E_{2n-1}+E_{2n}}{2}=\lambda_n+O\left(\epsilon^2 \right).
$$
The natural compactification of $\Gamma$ has two infinity points:
$$
\infty_{+}: \mu\sim-\lambda^{2N+1}, \ \  \infty_{-}: \mu\sim \lambda^{2N+1}.
$$
We have the following system of cuts (marked by the dashed lines in Figure~\ref{fig:basis1}): $[i\infty,E_0]$,  $[E_1,E_2]$,\ldots, $[E_{2N-1},E_{2N}]$, $[\overline{E_{2N}},\overline{E_{2N-1}}]$,  $[\overline{E_{2}},\overline{E_{1}}]$, 
$[\overline{E_{0}},-i\infty]$.  We use a slightly non-standard agreement: $\infty_+$ is located in Sheet~2 (dashed lines), and  $\infty_{-}$ is located in Sheet~1 (solid lines). 

To obtain convenient formulas, it is essential to choose a proper basis of cycles. In our text we use the one illustrated in Figure 7.

To calculate finite-gap solutions we need the periods of the basic holomorphic differentials and some meromorphic 
differentials, the vector of Riemann constants 
and the Abel transform of the Dirichlet spectrum. Let us calculate them to the leading order. In our text we use the following agreements:
\begin{enumerate}
\item The basic holomorphic differentials are normalized by:
\begin{equation}
\label{eq:omegas}
\big(\vec a_k\big)_j =  \oint\limits_{a_j}\omega_k = 2\pi i.
\end{equation}
In this normalization the real part of the Riemann matrix $B=(b_{jk})$
\begin{equation}
\label{eq:riemann0}
b_{jk}=\big(\vec b_k\big)_j =  \oint\limits_{b_j}\omega_k
\end{equation}
is \textbf{negative defined}.
\item In Sections~\ref{sec:sec5.1}, \ref{sec:sec5.2} we assume that the starting point for the Abel transform is the branch point  $E_0=i+O(\epsilon^2)$:
\begin{equation}
\label{eq:Abel}
\vec A(\gamma) = \vec A_{E_0}(\gamma) =  \left[\begin{array}{c} \int\limits_{E_0}^{\gamma} \omega_1 \\ \vdots \\ \int\limits_{E_0}^{\gamma} \omega_{2N} \end{array}\right] =
\int\limits_{E_0}^{\gamma} \vec\omega,  \ \ \mbox{where} \ \  \vec\omega =\left[\begin{array}{c} \omega_1 \\ \vdots \\ \omega_{2N} \end{array}\right], \ \ \gamma\in\Gamma.
\end{equation}
\end{enumerate}
\subsection{The Riemann matrix}
\label{sec:sec5.1}

The calculation of the diagonal elements to the leading order does not depend on $N$; therefore we can use the results of the paper \cite{GS1}:
\begin{equation}
\label{eq:riemann2}
b_{jj}= 2 \log\left[ \frac{\epsilon\sqrt{\hat\alpha_j\hat\beta_j}}{|4\sin(2\hat\phi_j) \cos(\hat\phi_j)|} \right]+O(\epsilon), \ \ 1\le j \le 2N.
\end{equation}
$$
\exp{\left(\frac{b_{jj}}{2}\right)}=
\frac{\epsilon\sqrt{\hat\alpha_j\hat\beta_j}}{|4\sin(2\hat\phi_j) \cos(\hat\phi_j)|}+O(\epsilon^2), \ \ 1\le j \le 2N,
$$
where $\hat\alpha_j,\hat\beta_j$ are defined in (\ref{def_hat_alpha_beta}). 
Up to $O(\epsilon^2)$ corrections, the off-diagonal elements do not depend on the perturbation, and coincide with the limiting values from \cite{ItsRybinSall}:
\begin{equation}
\label{eq:b_off_diag}
b_{jk} = 2\log \left|\frac{\sin\left(\frac{\hat\phi_j-\hat\phi_k}{2} \right)}{\cos\left(\frac{\hat\phi_j+\hat\phi_k}{2} \right) }\right| +O(\epsilon^2) , \ \ j\ne k, \ j,k=1,\ldots,2N.
\end{equation}

\subsection{The vector of Riemann constants}
\label{sec:sec5.2}

The calculation of the Riemann constant vector is rather standard (see, for example, \cite{Dubrovin2}), but we present it for completeness. It is based on the periodicity properties of the Riemann theta-functions (\ref{eq:theta-per})
\begin{align}
\theta(\vec A(\gamma)-{\vec C} + \vec a_k)  =  &\ \theta(\vec A(\gamma)-\vec C)&\\
\theta(\vec A(\gamma)-{\vec C} + \vec b_k)  =  &\ \theta(\vec A(\gamma)-\vec C)\exp\left(-\frac{1}{2}b_{kk}- A_k(\gamma) + C_k\right),&
\end{align}
where $\vec C$ is a generic vector and $C_k$ are its components. Therefore
\begin{equation}
\begin{split}
&\log\big[\theta(\vec A(\gamma)-C + \vec a_k) \big]  =  \ \log\big[\theta(\vec A(\gamma)-C)\big] + 2\pi i L_k, \ \  L_k\in \ZZ, \\
&\log\big[\theta(\vec A(\gamma)-C + \vec b_k)\big]   =  \\
&\hspace{2cm}= \log\big[\theta(\vec A(\gamma)-C)] -\frac{1}{2} b_{kk}- A_k(\gamma) + C_k + 2\pi i M_k, \ \ M_k\in \ZZ,\\
&\int_{a_k}d\log\big[\theta(\vec A(\gamma)-C) \big]= 2\pi i L_k,\\
&\int_{b_k}d\log\big[\theta(\vec A(\gamma)-C) \big]= -\frac{1}{2} b_{kk}- A_k(\mbox{starting point of }b_k) + C_k + 2\pi i M_k,\\
&d\log\big[\theta(\vec A(\gamma)-C+\vec a_k) \big] =d\log\big[\theta(\vec A(\gamma)-C) \big],\\
&d\log\big[\theta(\vec A(\gamma)-C+\vec b_k) \big] =d\log\big[\theta(\vec A(\gamma)-C) \big]- \omega_k.
\end{split}
\end{equation}

\begin{figure}[H]
\centering
\parbox{7cm}{\epsfysize=14cm\epsffile{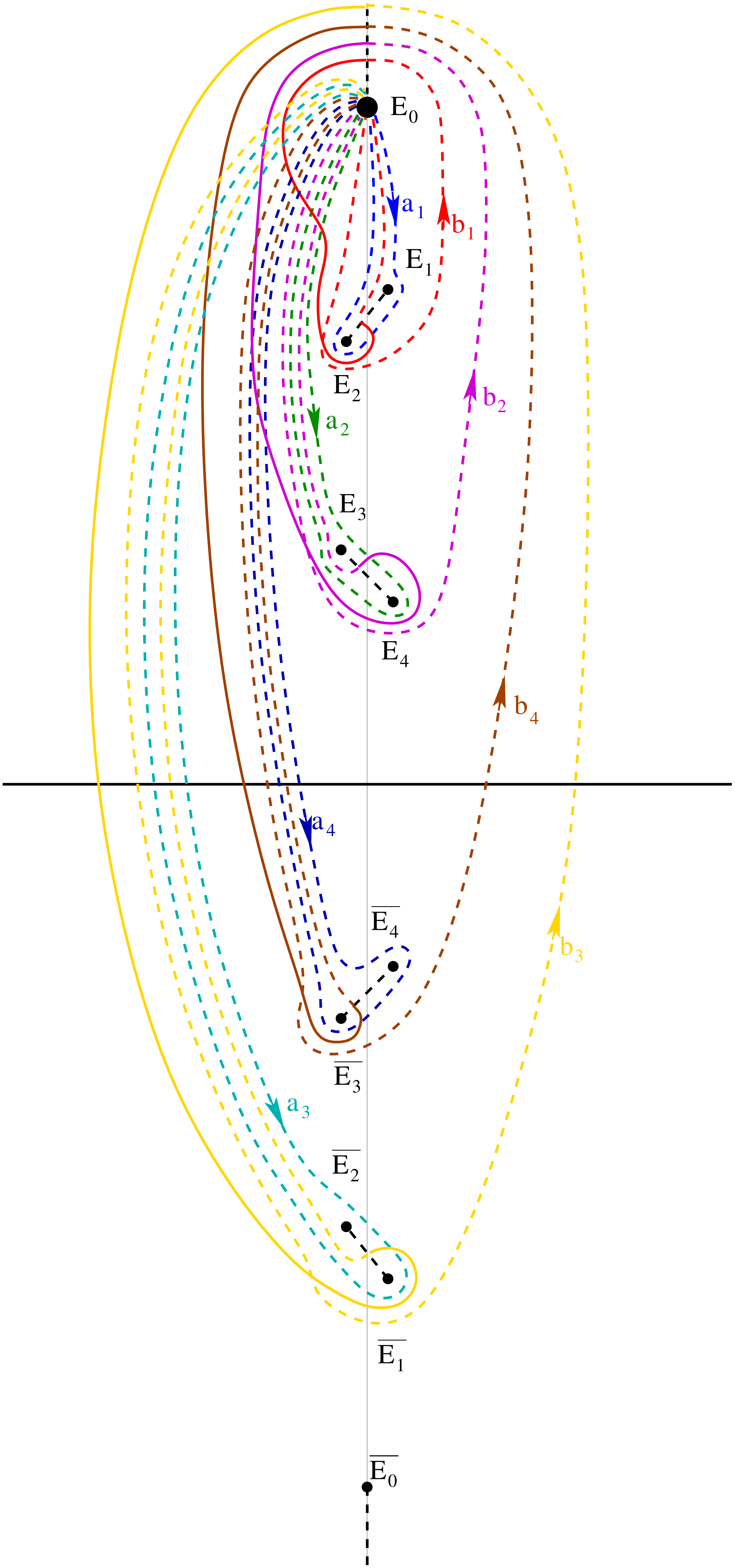}}%
\parbox{8.5cm}{\begin{center}\epsfxsize=3cm\epsffile{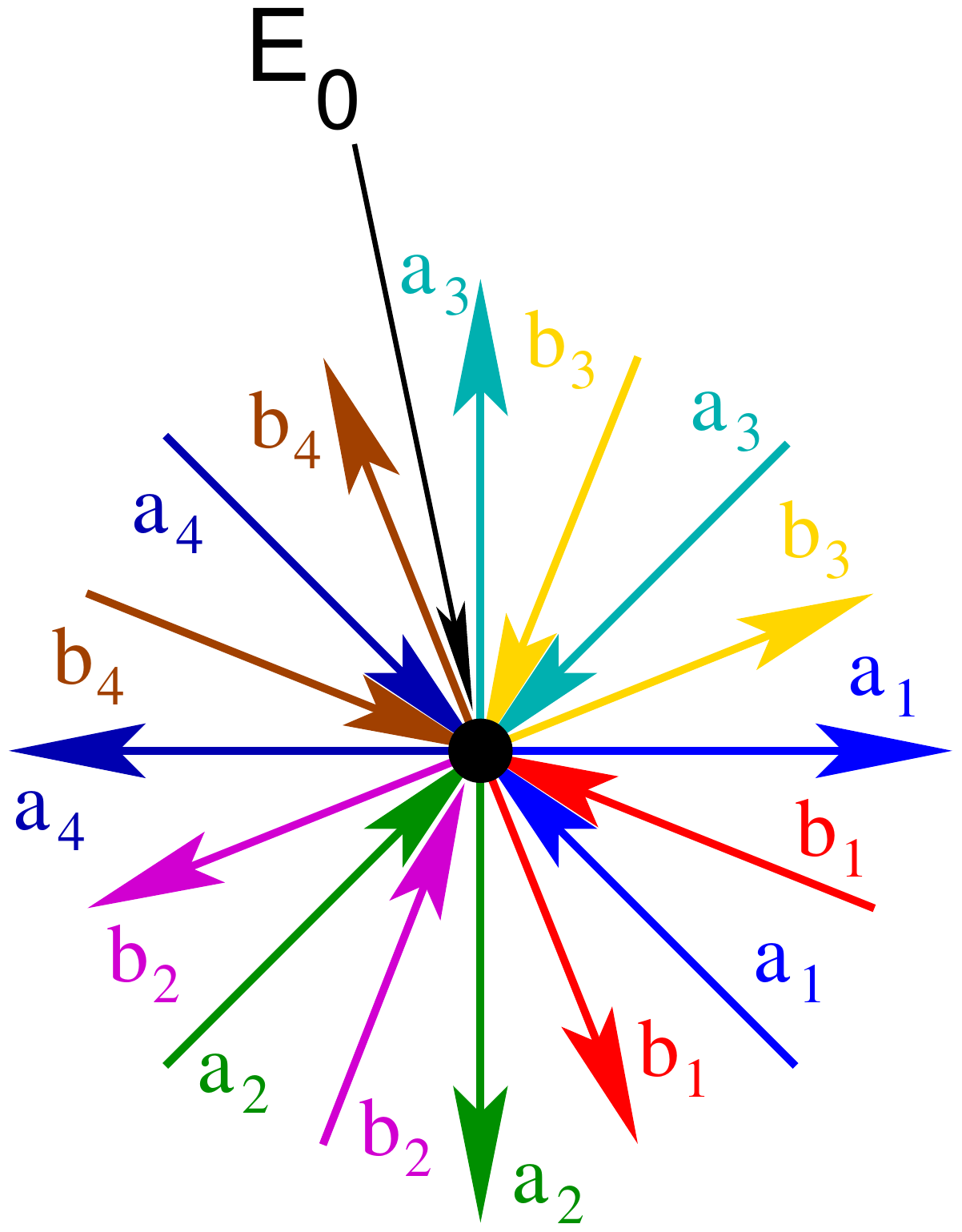} 

\vspace{2cm}  

\epsfxsize=8cm\epsffile{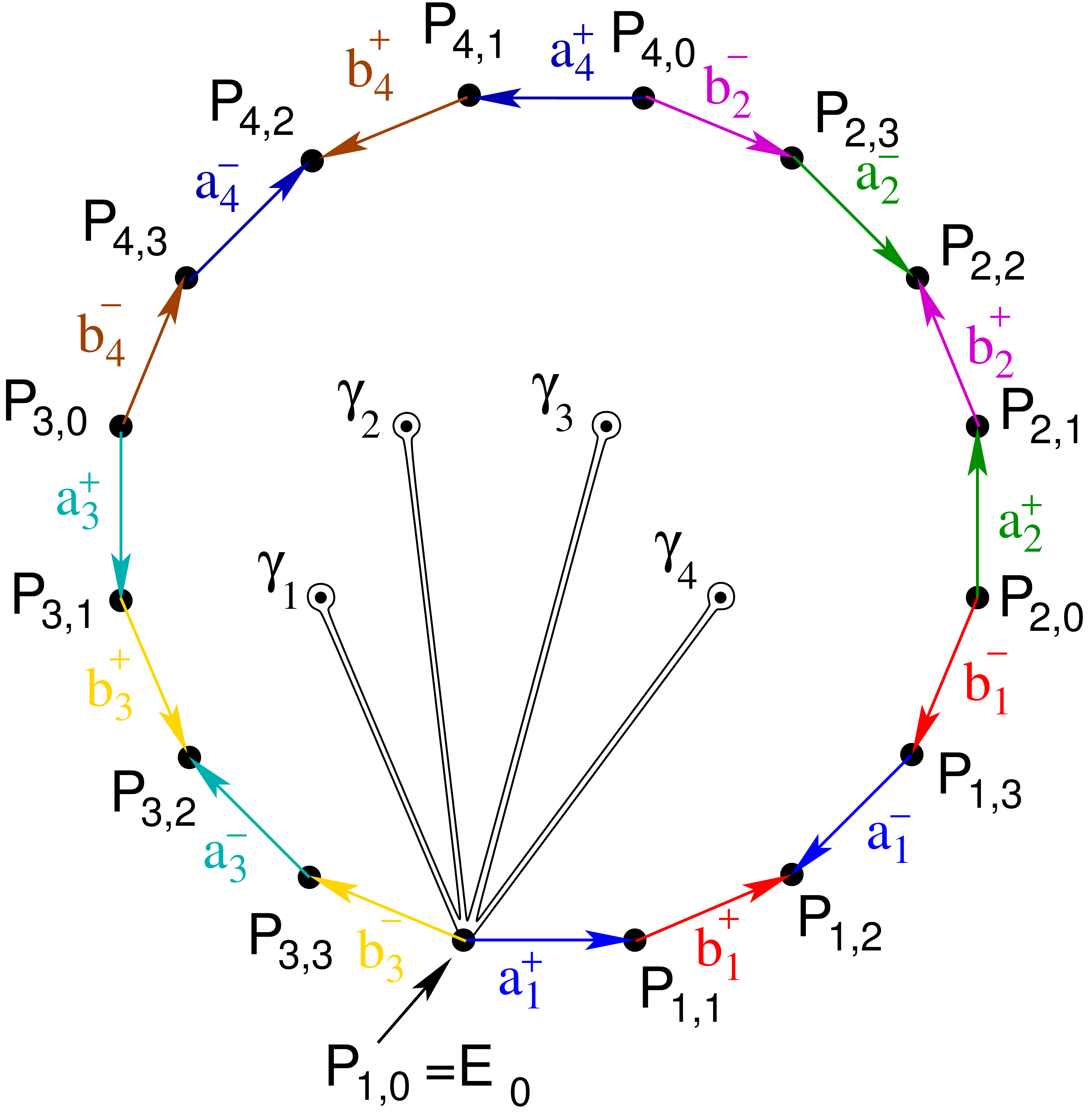} \end{center}}
\caption{\label{fig:8N-gone} On the left: the system of cuts on $\Gamma$ for $N=2$. On the right above: the order of cycles near the point $E_0$. On the right below: the standard $8N$-gone for $N=2$. The starting point is also connected with the zeroes $\gamma_1$, \ldots,  $\gamma_{2N}$  of the function $\theta(\vec A(\gamma)-\vec C)$ The order of the basic cycles along the $8N$-gone agrees with the order of the cycles in the figure above.}
\end{figure}

Consider the $8N$ gone obtained from $\Gamma$ by cutting along the cycles. To do such cutting it is important to have a non-intersecting system of cycles starting from the point $E_0$ (see Figure~\ref{fig:8N-gone}). 

Let us remark that

\begin{equation}
\begin{split}
\vec A(P_{k,0}) = \vec 0, \ \ \ \ \ \ \ & \vec A(P_{k,1}) = \vec a_k, \\
\vec A(P_{k,2}) = \vec a_k+ \vec b_k, \ \ \ \ \ \ & \vec A(P_{k,3}) = \vec b_k. \\
\end{split}
\end{equation}

Denote by $\DD=\gamma_1+\ldots+\gamma_{2N}$ the divisor of zeroes of the function $\theta(\vec A(\gamma)-\vec C)$ (since $\vec C$ to be generic, this function is not  identically equal to zero). Integrating the form  $\omega_j \log\big[\theta(\vec A(\gamma)-\vec C) \big]$ either along the paths connecting the point $E_0$ with the divisor points or along the $8N$-gone, we obtain:
$$
-2\pi i \vec A_j(\DD) = \int_{\partial\Gamma} \omega_j \log\big[\theta(\vec A(\gamma)-C) \big]
$$
$$
=\sum_{k=1}^{2N} \left[\int_{a^+_k}\omega_j(\gamma)\log\big[\theta(\vec A(\gamma)-C) \big]+
\int_{b^+_k} \omega_j \log\big[\theta(\vec A(\gamma)-C) \big]-\right.
$$
$$
\left. -\int_{a^-_k} \omega_j \log\big[\theta(\vec A(\gamma)-C) \big]-
\int_{b^-_k} \omega_j \log\big[\theta(\vec A(\gamma)-C) \big]\right]=
$$
$$
=\sum_{k=1}^{2N} \left[\int_{a^+_k} \omega_j \log\big[\theta(\vec A(\gamma)-C) \big]  \right.
$$
$$
-\int_{a^+_k} \omega_j \left[\log\big[\theta(\vec A(\gamma)-C) \big]  -\frac{1}{2} b_{kk}- A_k(\gamma) +  C_k + 2\pi i M_k \right]  
$$
$$
\left. -\int_{b^-_k} \omega_jd\log\big[\theta(\vec A(\gamma)-C) \big]  + 
\int_{b^-_k} \omega_j \big[\log\big[\theta(\vec A(\gamma)-C) \big]+2\pi i L_k  \big]  \right]  
$$
$$
=\sum_{k=1}^{2N} \left[\int_{a^+_k} \omega_j \left[\frac{1}{2} b_{kk}+ A_k(\gamma) - C_k - 2\pi i M_k \right] +
\int_{b^-_k} \omega_j 2\pi i L_k  \big]\right]   
$$
$$
=2\pi i \left[ \frac{b_{jj}}{2}- C_j - 2\pi i M_j +2 \pi i L_j + \frac{1}{2\pi i}\sum_{k=1}^{2N} \left[\int_{a^+_k} \omega_j A_k(\gamma) \right]  \right].
$$
Finally, modulo period, we obtain
$$
\vec A_j(\DD) =  -\frac{b_{jj}}{2} + C_j  -\frac{1}{2\pi i} \sum_{k=1}^{2N} \left[\int_{a^+_k} \omega_j A_k(\gamma) \right] = C_j- K_j,
$$
\begin{equation}
\label{eq:riemann_const2}
K_j = \frac{b_{jj}}{2} - \pi i + \frac{1}{2\pi i}\sum_{ k=1, k\ne j }^{2N} \int_{a^+_k} A_k(\gamma) \omega_j .
\end{equation}
In addition, if $j\ne k$, then
$$
\int_{a^+_k} A_j(\gamma) \omega_k = 2\pi i A_j(\mbox{one of the branch points of } a_k);
$$
therefore, at last,
\begin{equation}
\label{eq:riemann_const3}
K_j = \frac{b_{jj}}{2} -\pi i - \sum_{ k=1, k\ne j }^{2N} \left\{\begin{array}{ll} A_j(E_{2k}), & k\le N, \\
 A_j(\overline{E_{2(k-N)-1}} ), & k > N.  \end{array} \right.
\end{equation}

From a simple direct calculation it follows that 
\begin{equation}
\label{eq:abel_bp}
\begin{split}
&\vec A(E_{0}) = \vec 0,\\
&\vec A(E_{1}) = \frac{1}{2}\left[ -\vec b_1 \right],\\
&\vec A(E_{2}) = \frac{1}{2}\left[ -\vec b_1 +\vec a_1 \right],\\
&\vec A(E_{3}) = \frac{1}{2}\left[ -\vec b_2 +\vec a_1  \right],\\
&\vec A(E_{4}) = \frac{1}{2}\left[ -\vec b_2 +\vec a_1 + \vec a_2 \right],\\
\ldots\\
&\vec A(E_{2k-1}) = \frac{1}{2}\left[\sum\limits_{j=1}^{k-1} \vec a_j -\vec b_k  \right],\ \ k\le N, \\
&\vec A(E_{2k}) = \frac{1}{2}\left[\sum\limits_{j=1}^{k} \vec a_j - \vec b_k \right], \ \ k\le N,\\
&\vec A(\overline{E_{0}}) =\frac{1}{2}\left[\sum\limits_{j=1}^{2N} \vec a_j  \right] ,\\
&\vec A(\overline{E_{1}}) = \frac{1}{2}\left[\sum\limits_{j=1}^{2N} \vec a_j   - \vec b_{N+1}\right],\\
&\vec A(\overline{E_{2}}) = \frac{1}{2}\left[\sum\limits_{j=1}^{2N-1} \vec a_j   - \vec b_{N+1}\right],\\
\ldots\\
&\vec A(\overline{E_{2k-1}}) = \frac{1}{2}\left[\sum\limits_{j=1}^{2N-k+1} \vec a_j -\vec b_{N+k}  \right],\ \ k\le N, \\
&\vec A(\overline{E_{2k}}) = \frac{1}{2}\left[\sum\limits_{j=1}^{2N-k} \vec a_j - \vec b_{N+k} \right], \ \ k\le N.\\
\end{split}
\end{equation}
We see that the $a_k$ and $b_k$ parts of $\vec K$ can be calculated separately: $\vec K = \vec K_a + \vec K_b$. 
For $\vec K_b$ we immediately obtain:
$$
\vec K_b = \frac{1}{2}\sum\limits_{j=1}^{2N} \vec b_j.
$$
Let us calculate $\vec K_a$. Modulo $b_k$ we have:
$$
\frac{1}{2} = \left\{\begin{array}{ll}  A_j(E_{2j}), & j \le N ,  \\   A_j(\overline{E_{2(j-N)-1}}), & j>N; 
\end{array}\right.
$$
therefore, modulo $b_k$, 
$$
(K_a)_j = - \sum_{ k=1 }^{2N} \left\{\begin{array}{ll} A_j(E_{2k}), & k\le N ,\\
 A_j(\overline{E_{2(k-N)-1}} ), & k > N,  \end{array} \right.
$$
and
$$
\vec K_a =\pi i\left[\begin{array}{c} 0 \\ 1 \\ 0 \\ \vdots \\ 1 \\ 0 \\ \hline  
1 \\ 0 \\ 1  \\ \vdots\\ 0 \\ 1  \end{array}\right], \ \ 
N \ \mbox{odd}; \ \ \ 
\vec K_a =\pi i \left[\begin{array}{c} 0 \\ 1 \\ 0 \\ \vdots \\ 1 \\ 0 \\ 1 \\ \hline 
1 \\ 0 \\ 1 \\ \vdots \\0 \\ 1 \\ 0 \end{array}\right], \ \ 
N \ \mbox{even}. 
$$

Finally, we obtain:
\begin{equation}
\label{eq:vec_K}
\vec K = \sum\limits_{k=1}^N \big[ \vec A(E_{2k})+  \vec A(\overline{E_{2k-1}})\big].
\end{equation}

Here we used the following argument. Each cycle $a_k$ consists of three parts: a path ${\cal P}_{k,1}$ connecting $E_0$ 
to the point $E_{2k-1}$ if $k\le N$, or to the point $\overline{E_{2(k-N)}}$ if $k>N$, a closed contour ${\cal P}_{k,2}$ 
going about either the interval $[E_{2k-1},E_{2k}]$ or $[\overline{E_{2(k-N)}},\overline{E_{2(k-N)-1}}]$ respectively, and the
path  ${\cal P}_{k,1}$ with the opposite orientation.

Assume that $k\ne j$. Then the function $A_j(\gamma)$ has the same values when we go forward and back along  
${\cal P}_{k,1}$; therefore this part of the integral vanishes: 
$$
\int\limits_{{\cal P}_{k,2}} A_j(\gamma)\omega_k(\gamma) = 
\int\limits_{{\cal P}_{k,2}} A_j(\mbox{BP}) \omega_k(\gamma) + 
\int\limits_{{\cal P}_{k,2}} [ A_j(\gamma)- A_j(\mbox{BP} )] \omega_k(\gamma).
$$
The function $ A_j(\gamma)- A_j(\mbox{BP})$ takes opposite values on the opposite sides of the cycle around the cut; 
therefore the second integral is equal to zero, and
$$
\int\limits_{a_k} A_j(\gamma)\omega_k(\gamma) =2\pi i A_j(\mbox{BP}),
$$
where BP denotes one of the branch points $E_{2k-1}$, $E_{2k}$ for $k\le N$, or one of the branch points 
$\overline{E_{2(k-N)}}$, $\overline{E_{2(k-N)-1}}$ for $k > N$. 
Let us also calculate 
$$
\int\limits_{a_k} A_k(\gamma)\omega_k(\gamma) = \frac{1}{2}\int\limits_{a_k} d\big(A_k^2(\gamma)\big)^2
$$
$$
=\frac{1}{2}\big[A_k^2(\mbox{final point of the path})-A_k^2(\mbox{starting point of the path})  \big]= \frac{(2\pi i)^2}{2}
$$

Formula (\ref{eq:vec_K}) admits the following interpretation. Assume that a divisor $\DD=\gamma_1+\ldots+\gamma_{2N}$ of degree $2N$ has the following structure: the point $\gamma_j$ is located near the point $\hat\lambda_j$. Then formula~(\ref{eq:pot1_1}) admits the following simplification:
\begin{equation}
\label{eq:K_simpl}
\vec A(\DD) + \vec K = \sum\limits_{j=1}^N \int\limits_{E_{2j}}^{\gamma_j} \vec \omega+  \sum\limits_{j=1}^N \int\limits_{\overline{E_{2j-1}}}^{\gamma_{j+N}} \vec\omega. 
\end{equation}
\textbf{Agreement.} In the remaining part of our paper we use the following notations: if a divisor point $\gamma_j$ is located close enough to the point $\hat\lambda_j$, then we calculate its Abel transform starting from the corresponding branch point:
\begin{equation}
\label{eq:abel_redef}
\vec A(\gamma_j)=\vec A_{E_{2j}}(\gamma_j), \ \  \vec A(\gamma_{j+N})= \vec A_{\overline{E_{2j-1}}}(\gamma_{j+N}), \ \ j=1,\ldots,N;
\end{equation}
therefore we can \textbf{redefine} the Abel transform of the divisor by: 
\begin{equation}
\label{eq:abel_redef2}
\vec A(\DD) = \sum\limits_{j=1}^{N} \bigg[\vec A_{E_{2j}}(\gamma_j)+ \vec A_{\overline{E_{2j-1}}}(\gamma_{j+N}) \bigg].
\end{equation}
Formula (\ref{eq:K_simpl}) means that, if in the theta-functional formulas, the Abel transform of the divisor is calculated using the Agreement (\ref{eq:abel_redef2}), then the vector of Riemann constants $\vec K$ is identically zero:
\begin{equation}
\label{eq:vec_K0}
\vec K = \left[\begin{array}{c} 0 \\ \vdots \\ 0 \end{array}\right].
\end{equation}
This simplification was not used in \cite{GS1}.

\subsection{Periods of some meromorphic differentials}
\label{sec:sec5.3}

The formulas for finite-gap solutions include the following meromorphic differentials:
\begin{enumerate}
\item A 3-rd kind meromorphic differential $\Omega$ with zero $a$-periods and first order poles at $\infty_+$,  
$\infty_-$:
\begin{align*}
\Omega &= -\frac{d\lambda}{\lambda}+O(1),& &\mbox{at} \ \ \infty_+ ,\\
\Omega &= \frac{d\lambda}{\lambda}+O(1),& &\mbox{at} \ \ \infty_- ,\\ 
\oint\limits_{a_j}\Omega&=0,& & j=1,\ldots,2N.
\end{align*}    
From the Riemann bilinear relations (see \cite{Spr1957}), it follows that:
$$
A_j(\infty_-)-A_j(\infty_+)=Z_j=\oint\limits_{b_j}\Omega.
$$
A sufficiently standard calculation (for $N=1$ the details can be found in \cite{GS1})) implies:
\begin{equation}
\label{eq:omega_per1}
\vec A(\infty_-) = - \vec A(\infty_+), \ \ 
\big(\vec A(\infty_+)\big)_j = \frac{\pi i}{2} +  i \hat\phi_j  + O(\epsilon^2).
\end{equation}
\item The differential of the multivalued quasimomentum function $p(\gamma)$ is the 2-nd kind meromorphic differential $dp$ (quasimimentum differentials) such that
\begin{align*}
dp &= -d\lambda +O(1),& &\mbox{at} \ \ \infty_+ , \\
dp &=  d\lambda +O(1),& &\mbox{at} \ \ \infty_- ,\\ 
\oint\limits_{a_j}dp &=0,&  j=1,\ldots,2N.
\end{align*} 
It is convenient to use the Riemann bilinear relations:
$$
\oint\limits_{b_j}dp=- \left[\res\nolimits_{\infty_+} [p\,\omega_j]+ \res\nolimits_{\infty_-} [p\, \omega_j]  \right].
$$
Again, a sufficiently standard calculation (for $N=1$ see \cite{GS1}) implies:
\begin{equation}
\label{eq:dp_per1}
\oint\limits_{b_j}dp=2\cos(\hat\phi_n) +O(\epsilon^2).
\end{equation}
Taking into account that 
\begin{equation}
\label{eq:U1}
\left(\vec U_1 \right)_j = - i\oint\limits_{b_j}dp,
\end{equation}
we immediately obatin
\begin{equation}
\label{eq:U1bis}
\left(\vec U_1 \right)_j=-2i\cos(\hat\phi_j) +O(\epsilon^2).
\end{equation}
\begin{remark}
If the finite-gap approximation is constructed using some periodicity preserving technique, for example the isoperiodic deformation from \cite{GrSch}, then, instead of (\ref{eq:dp_per1}), we have an \textbf{exact} relation:
\begin{equation}
\label{eq:dp_per2}
\oint\limits_{a_j}dp=2\cos(\hat\phi_n).
\end{equation}
Otherwise, if we close the gaps corresponding to the stable modes ``naively'', the spatial periodicity of the leading order solution  breaks at $O(\epsilon^2)$.
\end{remark}
\item A 2-nd kind meromorphic differential $dq$ such that
\begin{align*}
dq &= -d(\lambda)^2 +O(1),& &\mbox{at} \ \ \infty_+,\\
dq &=  d(\lambda)^2 +O(1),& &\mbox{at} \ \ \infty_-,\\ 
\oint\limits_{a_j}dq &=0,&  j=1,\ldots,2N.
\end{align*} 
Again, using the Riemann bilinear relations:
$$
\oint\limits_{b_j}dq=- \left[\res\nolimits_{\infty_+} [q\, \omega_j]+ \res\nolimits_{\infty_-} [q\, \omega_j]  \right],
$$
we obtain (for $N=1$ see \cite{GS3})
\begin{equation}
\label{eq:dq_per1}
\oint\limits_{b_j}dq= i\sin(2\hat\phi_j) +O(\epsilon^2).
\end{equation}
Taking into account that 
\begin{equation}
\label{eq:U2}
\left(\vec U_2 \right)_j = - 2i\oint\limits_{b_j}dq,
\end{equation}
we immediately obatin
\begin{equation}
\label{eq:U2bis}
\left(\vec U_2 \right)_j=2\sin(2\hat\phi_j) +O(\epsilon^2).
\end{equation}
\end{enumerate}

\subsection{Abel transform of the divisor}
\label{sec:sec5.4}

In this Section we calculate the Abel transform of divisor points assuming normalization (\ref{eq:abel_redef}) up to $O(\epsilon)$ correction.

First of all, under this assumption, 
$$
A_k(\hat\gamma_j) = O(\epsilon), \ \ k\ne j.
$$
and we will neglect it.

Let us calculate $A_j(\hat\gamma_j)$. 

Near the point $\hat\lambda_j$ we have:
$$
p(\gamma) = \left\{ \begin{array}{ll} p(E_{2j})+\frac{\hat\lambda_j}{\hat\mu_j}\nu(\gamma)+O(\epsilon^2), & j\le N, \\ 
p(\overline{E_{2j-2N-1}})+\frac{\hat\lambda_j}{\hat\mu_j}\nu(\gamma)+O(\epsilon^2),  & j > N.
     \end{array} \right. 
$$
where
$$
\nu^2=\left\{ \begin{array}{ll} (\lambda-E_{2j-1}) (\lambda-E_{2j}), &  j\le N, \\ 
(\lambda-\overline{E_{2j-2N-1}}) (\lambda-\overline{E_{2j-2N}}), & j>N.
      \end{array} \right.
$$
We also have:
$$
\omega_j = \frac{d\lambda}{\nu}+O(\epsilon) = d\log\left[\lambda(\gamma)-\hat\lambda_j+\nu(\gamma)   \right]+O(\epsilon)
$$
and 
$$
A_j(\hat\gamma_j) = \log\left[\frac{\lambda(\hat\gamma_j)-\hat\lambda_j+\nu(\hat\gamma_j)}{E_s-\hat\lambda_j} \right],
+O(\epsilon)
$$
where
$$
E_s=\left\{ \begin{array}{ll} E_{2j}, & j\le N, \\ \overline{E_{2j-2N-1}}, & j>N.  \end{array}\right.
$$
We have, up to $O(\epsilon^2)$ corrections:
$$
E_s-\hat\lambda_j=\left\{ \begin{array}{ll} \frac{\epsilon}{2\hat\lambda_j} \sqrt{\alpha_j\beta_j}, & j\le N,  \\
                            -\frac{\epsilon}{2\hat\lambda_{j} }\sqrt{\hat\alpha_{j}\hat\beta_{j}}, & j > N.  
       \end{array}\right.
$$

Here we asssume:
$$
\Re  \sqrt{\alpha_j\beta_j}\ge 0, \ \ \ \ \sqrt{\hat\alpha_{j+N}\hat\beta_{j+N}} =\overline{\sqrt{\alpha_j\beta_j}}.
$$
Using (\ref{eq:div_alpha1}), (\ref{eq:div_alpha2}), we obtain:
$$
\lambda(\hat\gamma_j)-\hat\lambda_j+\nu(\hat\gamma_j)=\left\{ \begin{array}{ll} 
e^{i\hat\phi_j}\frac{\epsilon}{2\hat\lambda_j} \hat\alpha_j+O(\epsilon^2), & j\le N , \\
-e^{i\hat\phi_j} \frac{\epsilon}{2\hat\lambda_{j}} \hat\alpha_{j}+O(\epsilon^2),  & j > N, \end{array}\right.
$$
and, finally,
$$
A_j(\hat\gamma_j) = 
\log\left[\frac{\hat\alpha_j}{\sqrt{\hat\alpha_j\hat\beta_j}} \right]+i\hat\phi_j + O(\epsilon).
$$

\section{Theta-functional solutions}
\label{sec:sec6}

The finite-gap leading order solution of the Cauchy problem of the anomalous waves is provided by formula (\ref{eq:pot1_1}), with
\begin{equation}
\label{eq:pot1_params}
{\cal C} = \frac{\theta(\vec A(\infty_+)-\vec A(\DD)|B)}
{\theta(\vec A(\infty_-)-\vec A(\DD)|B)}  u(0,0)(1+O(\epsilon^2)),\ \ {\cal U} = O(\epsilon^2), \ \ {\cal V} = 2i + O(\epsilon^2).
\end{equation}
Taking into account Agreement~(\ref{eq:abel_redef2}), and using (\ref{eq:vec_K0}), (\ref{eq:omega_per1}),  (\ref{eq:U1bis}), (\ref{eq:U2bis}), we obtain the following approximation for the arguments of the theta-function:
\begin{align}
\label{eq:pot2_params}
&\vec A(\infty_{+}) -\vec U_1 x - \vec U_2 t -\vec A(\DD) -\vec K = \vec z_{-}(x,t)   + O(\epsilon), \nonumber\\
&\vec A(\infty_{-}) -\vec U_1 x - \vec U_2 t -\vec A(\DD) -\vec K  = \vec z_{+}(x,t) + O(\epsilon), \nonumber\\
&\big(\vec z_{-}(x,t)\big)_j =\frac{i\pi}{2}  - \log\left[\frac{\hat\alpha_j}{\sqrt{\hat\alpha_j\hat\beta_j}} \right]  + 2i\cos(\hat\phi_j) x -2\sin(2\hat\phi_j) t \\
&\big(\vec z_{+}(x,t)\big)_j = \big(\vec z_{-}(x,t)\big)_j -\pi i - 2 i \hat\phi_j,
\end{align}
and, finally, we obtain the leading order solution 
\begin{equation}
\label{eq:pot1}
u(x,t)=\exp(2it)\cdot \frac{\theta( \vec z_{+}(x,t)|B)}{\theta(\vec z_{-}(x,t) |B)}\cdot (1+O(\epsilon)).
\end{equation}
in terms of the genus $2N$ $\theta$-functions.

\section{The solution of the Cauchy problem in terms of elementary functions}
\label{sec:sec7}

Formula (\ref{eq:pot1}) provides the solution up to $O(\epsilon)$ corrections. Therefore it is enough to sum the exponents in the theta-function over the elementary hypercube in $\RR^{2N}$ containing the trajectory point $-\vec w(t)$, where  $\vec w(t)$ is defined in (\ref{eq:vec_w}), and we obtain:
$$
\theta(\vec{z}_{\pm}(x,t)|B) = \tilde\theta(\vec{z}_{\pm}(x,t)|B) (1+O(\epsilon)),
$$
\begin{equation}
\label{eq:theta-hyper0}
\tilde\theta(\vec{z}_{\pm} (x, t)|B)= \!\!\!\!\!\!\!\!\!\!\!\!
\sum\limits_{\begin{array}{c}n_j^{\mbox{min}}(t) \le n_j \le n_j^{\mbox{max}}(t)\\ j=1,2,\ldots,2N\end{array}}
 \!\!\!\!\!\!\!\!\!\!\!\!
\exp{\left[\frac{1}{2} \sum\limits_{l=1}^{2N} \sum\limits_{s=1}^{2N} b_{ls}n_ln_s +  \sum\limits_{l=1}^{2N}  n_l (\vec{z}_{\pm}(x,t))_l  \right]},
\end{equation}
$$
n_j^{\mbox{min}}(t) = -\lfloor w_j(t) +1 \rfloor, \ \ n_j^{\mbox{max}}(t) = -\lfloor w_j(t) \rfloor.
$$
Here the floor function $\lfloor x \rfloor$ denotes the largest integer less of equal to $x$.

If we are interested in constructing the solution up to order $|\epsilon|^{p}$ corrections, $0<p<1$, for generic $t$, only a subset of vertices of that hypercube contributes. Therefore formula~(\ref{eq:theta-hyper0}) admits a further simplification.

To estimate the summands in the $\theta$-functions expansions,  we use the following identity:
$$
\Re\left(\sum\limits_{l=1}^{2N} \sum\limits_{s=1}^{2N} b_{ls}n_ln_s +  2\, \sum\limits_{l=1}^{2N}  n_l (  z_{\pm}(x,t))_l \right)=
\sum\limits_{l=1}^{2N} \sum\limits_{s=1}^{2N} \Re(b_{ls})n_ln_s +  2\, \sum\limits_{l=1}^{2N}  n_l \Re( z_{\pm}(x,t))_l  =
$$
\begin{equation}
\label{eq:distance}
=
\sum\limits_{l=1}^{2N} \sum\limits_{s=1}^{2N} \Re(b_{ls})( n_l+ w_l)( n_s+ w_s) - \sum\limits_{l=1}^{2N} \sum\limits_{s=1}^{2N} \Re(b_{ls}) w_l w_s.  
\end{equation}
Consider the metric on $\RR^{2N}$ with the metric tensor $g_{kl}=\Re b_{kl}$. Denote the distance between the point $-\tilde w(t)$ and the closest vertex of this hypercube by $d_0$. Relation (\ref{eq:distance}) means that the real parts of the arguments of the exponent in the $\theta$-series are equal, up to a common constant (the second term in the right hand side of (\ref{eq:distance})), to the distance between the point $-\tilde w$ and the corresponding lattice point. Therefore it is sufficient to keep in (\ref{eq:theta-hyper0}) only the vertices $\vec n$ such that the distance between  $-\tilde w(t)$ and $\vec n$ is smaller than a critical distance:
$$
d_{\mbox{cr}} = d_0 +\frac{p}{\pi}\left|\log(\epsilon) \right|,
$$
and in formula (\ref{eq:theta-hyper0}) we sum only over the corresponding $2{\cal N}(t)$ dimensional sub-cube of the full hypercube, ${\cal N}(t)\le N$, where $2{\cal N}(t)$ is the number of $j$'s such that $\tilde n_j^{\mbox{min}}(t)< \tilde n_j^{\mbox{max}}(t)$, with:
$$
\tilde n_j^{\mbox{min}}=-\lfloor w_j(t) +1/2 + p/2 \rfloor, \ \  \tilde n_j^{\mbox{max}}(t) = -\lceil w_j(t) -1/2 -p/2 \rceil.
$$
Here the ceiling function $\lceil x \rceil$ denotes the smallest integer greater or equal to $x$. It is easy to verify that this number is always even. Restricting the summation to this subset of vertices means that we approximate the NLS solution by the exact ${\cal N}(t)$-soliton solution of Akhmediev type (see formula (\ref{eq:Akh_CS})), whose parameters are written down through the Cauchy data in terms of elementary functions.

For numerical simulations, the representation  (\ref{eq:theta-hyper0}) is not very convenient because it involves ratios of exponentials with big arguments. To avoid this problem, one can use the periodicity properties of the theta-functions (\ref{eq:theta-per}) and shift the arguments to the basic elementary cell:
$$
(\tilde z_{\pm})_j(x,t) = (\vec{z}_{\pm}(x,t))_j - \sum\limits_k b_{jk} \lfloor w_j(t) \rfloor.
$$
Introducing also:
$$
\tilde w_j(t) = w_j(t) - \lfloor w_j(t) \rfloor,
$$
then equation (\ref{eq:pot1}) becomes 
\begin{equation}
\label{eq:pot11}
u(x,t)=\exp\left(2it + 2 i \Phi  \right)  
\frac{\theta(\tilde z_{+}(x,t)|B)}
{\theta(\tilde z_{-}(x,t)|B)}\times(1+O(\epsilon)),
\end{equation}
where
$$
\Phi = \sum\limits_{j=1}^{2N}  (\frac{\pi}{2} + \hat\phi_j)  \lfloor w_j(t) \rfloor .
$$
If the argument $z$ belongs to the basic elementary cell, to obtain the $O(\epsilon)$ approximation, we sum over the exponentials of the fundamental hypercube:
\begin{equation}
\label{eq:theta-hyper1}
\tilde\theta(z|B)= \sum\limits_{\begin{array}{c}n_j\in\{-1,0\}\\ j=1,2,\ldots,2N\end{array}}
\exp{\left[ \pi i \left(\sum\limits_{l=1}^{2N} \sum\limits_{s=1}^{2N} b_{ls}n_ln_s +  2\, \sum\limits_{l=1}^{2N}  n_l z_l \right) \right]}.
\end{equation}
Equivalently, shifting the hypercube by $1/2$ in all directions ($\hat n_j =2n_j+1$), one can write:
$$
u(x,t)=\exp\left(2it + 2 i \Phi  \right) \frac{\hat\theta(\hat z_{+}(x,t)|B)}
{\hat\theta(\hat z_{-}(x,t)|B)}  (1+O(\epsilon)),
$$
\begin{equation}
\label{eq:theta-hyper2}
\hat\theta(\hat z|B) = \sum\limits_{\begin{array}{c}{\hat n_j\in\{-1,1\}} \\ j=1,2,\ldots,2N\end{array}}
\exp{\left[ \sum\limits_{\begin{array}{c}l,s=1\\ l\ne s\end{array}}^{2N}\log \left|\frac{\sin\left(\frac{\hat\phi_l-\hat\phi_s}{2} \right)}{\cos\left(\frac{\hat\phi_l+\hat\phi_s}{2} \right) }\right| \frac{\hat n_l\hat n_s}{4} + \sum\limits_{l=1}^{2N}  \frac{\hat n_l \hat z_l}{2} - \frac{1}{2} \sum\limits_{l=1}^{2N} \hat z_j \right]},
\end{equation}
where
$$
\hat z_j = \tilde z_j -\frac{1}{2} \sum\limits_k b_{jk}.
$$

If the number of unstable modes is not too large (see Remark 11), the off-diagonal terms in $B$ could be omitted in the above rule for selecting the subset of  hypercube vertices providing the main contribution (but, of course, the off-diagonal terms should be kept in the arguments of the exponents in (\ref{eq:theta-hyper2})). 

Then we obtain the following rule:
\begin{enumerate}
\item If $\tilde w_j < \frac{1-p}{2}$ we keep only the terms with $n_j=1$.
\item If $\frac{1-p}{2}\le \tilde w_j \le \frac{1+p}{2}$ we keep the terms with $n_j=1$ and with $n_j=-1$.
\item If $\tilde w_j > \frac{1+p}{2}$ we keep only the terms with $n_j=-1$.
\end{enumerate}
Therefore, for each generic $t$, we have a summation over the vertices of a hypercube of dimension $2{\cal N}(t)$, ${\cal N}(t)\le N$. In our approximation the off-diagonal terms of matrix $B$ do not depend on $\epsilon$; therefore, in each time interval $I$, the approximation function does not depend on $\epsilon$, and in each time interval we approximate the solutions by an \textbf{exact} NLS solution, the ${\cal N}(I)$-soliton solution of Akhmediev type (\ref{sol_approx})-(\ref{phases}) (see Figures \ref{fig:3D_L20}).

\begin{figure}[H]
\centering
\includegraphics[width=14cm]{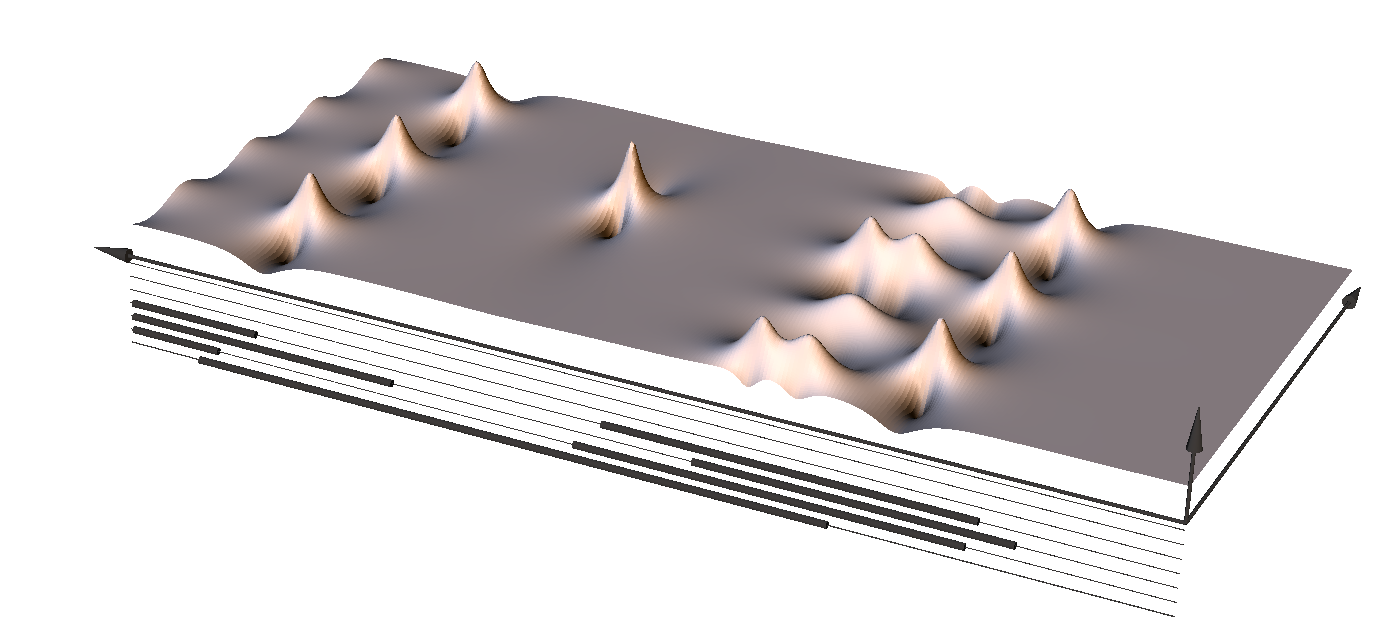} \\
\caption{\label{fig:3D_FG} The graph of the finite-gap approximation of $|u(x,t)|$, with $L=14$ (4 unstable modes), $0\le t \le 30$, $c_1 = 0.5$,  $c_{-1} = 0.3+0.3i$, $c_{2} = 0.5$, $c_{-2} = -0.03+0.03i$, $c_3 = 0.3$, $c_{-3} = 0.2+0.3i$, $c_4 = 0.3$, 
$c_{-4} = -0.3+0.03i$, $\epsilon=10^{-6}$. The graph below shows how the number ${\cal N}(t)$ depends on $t$. The graphs obtained by numerical integration, by applying the full-hypercube finite-gap approximation (\ref{eq:theta-hyper1}), and the graph obtained using the relevant vertices only are identical pixel-to-pixel; therefore we show only one of them.  We see that, after reaching the first nonlinear stage of MI, we do not return to the pure background state in the time interval studied. The appearance time of single peak in the middle of the graph is $T_{max}=18.76900$ from the numerical simulations,  $T_{max}=18.76906$ from the calculation using formula (\ref{eq:theta-hyper1}), and $T_{max}=18.76901$ from the analytic estimate (\ref{eq:tmax}).}
\end{figure}

\begin{figure}[H]
\centering
\includegraphics[width=10cm]{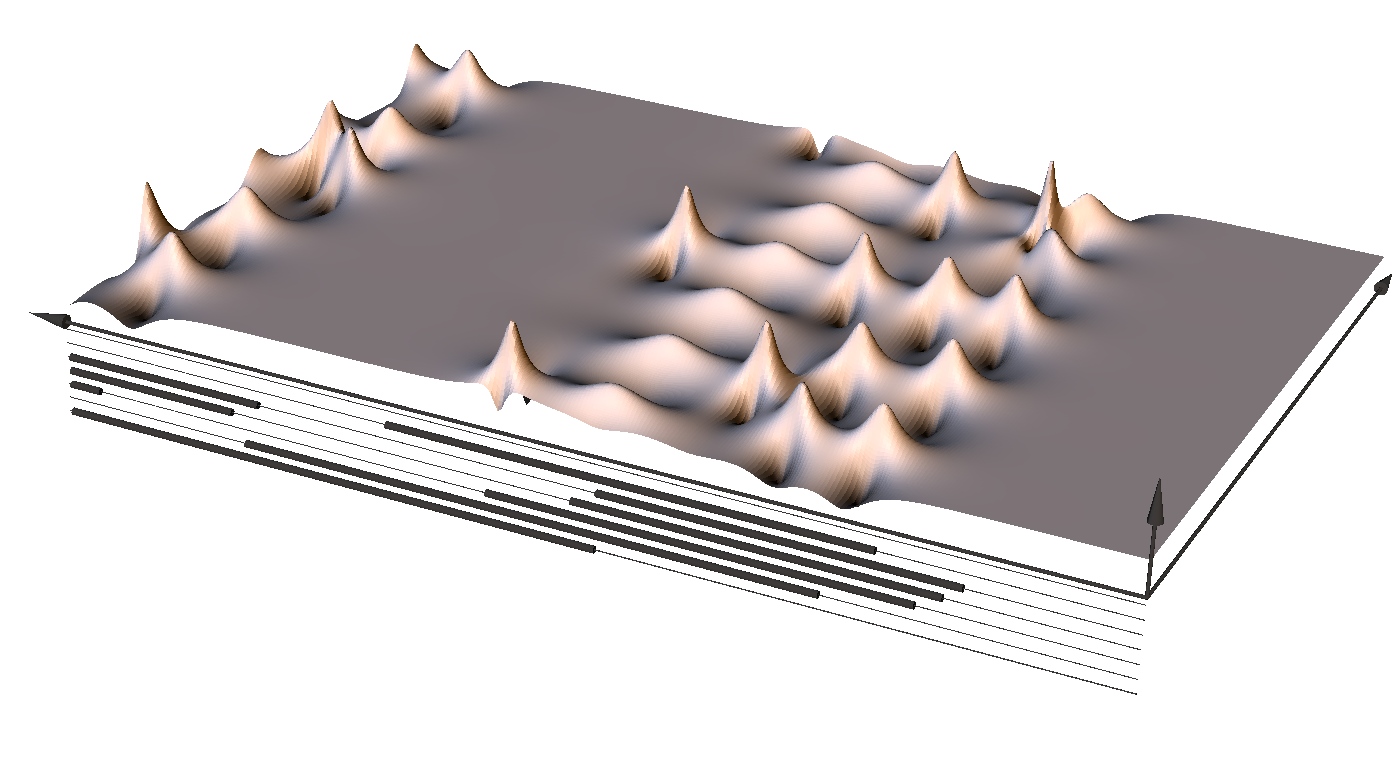}\\
\vspace{-0.5cm}
\includegraphics[width=10cm]{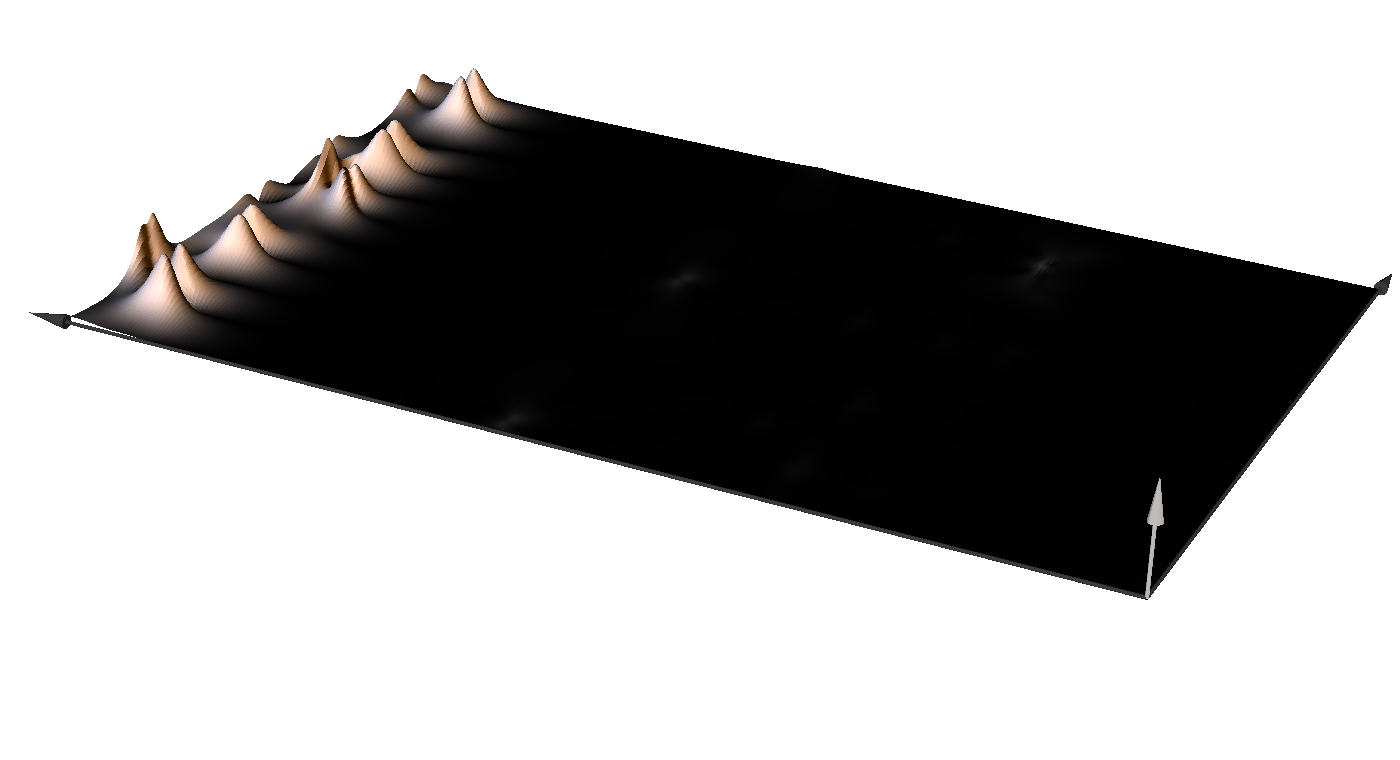}\\
\vspace{-1.2cm}
\includegraphics[width=10cm]{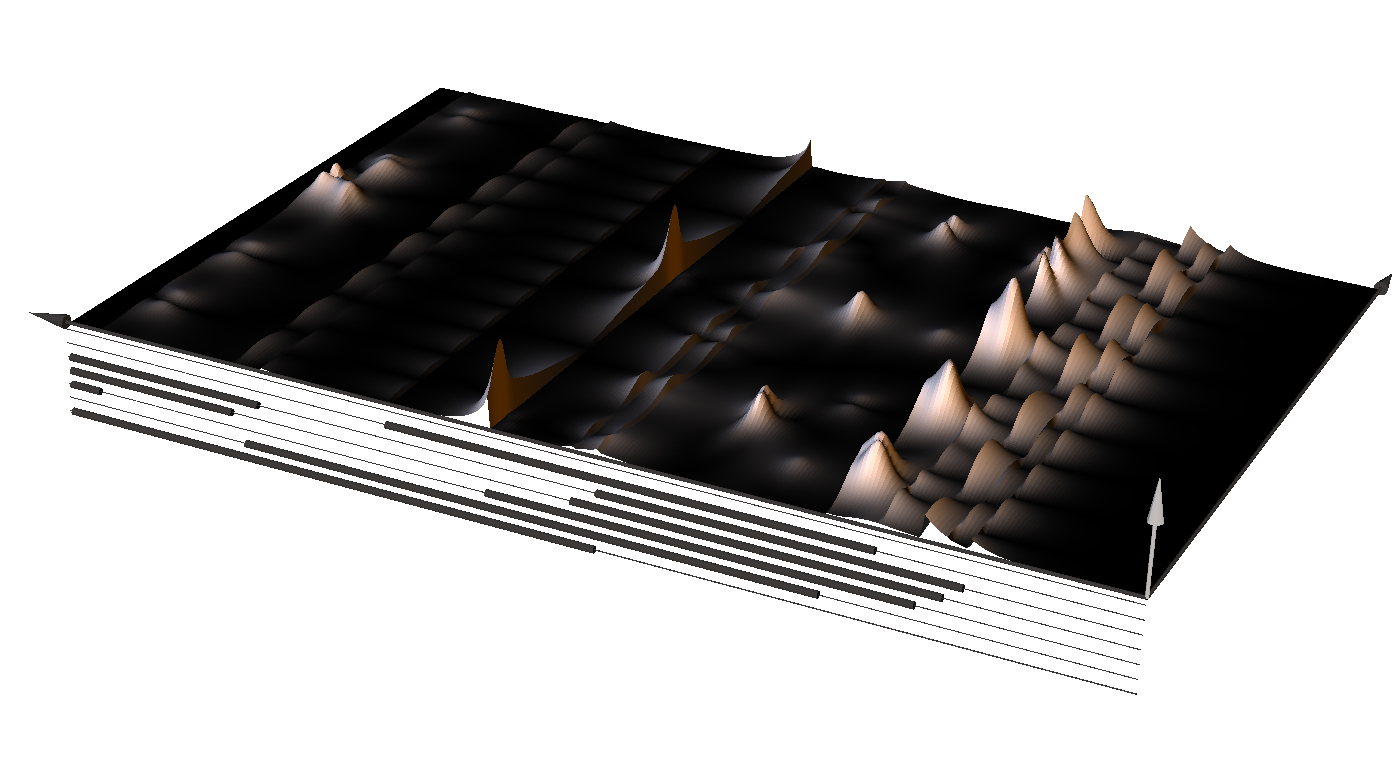}
\caption{\label{fig:3D_L20} Here we compare the absolute value of the numerical solution, the full-hypercube finite-gap approximation and the approximation involving the relevant vertices only. Here $L=20$ (6 unstable modes), $0\le t \le 30$, $c_1 = 0.5$,  $c_{-1} = 0.3+0.3i$, $c_{2} = 0.5$, $c_{-2} = -0.03+0.03i$, $c_3 = 0.3$, $c_{-3} = 0.2+0.3i$, $c_4 = 0.3$, $c_{-4} = -0.3+0.03i$, $c_5 = 0.3$, $c_{-5} = 0.2+0.3i$, $c_6 = 0.3$, $c_{-6} = -0.3+0.03i$, $\epsilon=10^{-6}$, $p=1/2$. The figure above shows the graph of $|u(x,t)|$ obtained using the analytic formula (\ref{eq:pot11}), coinciding pixel-to-pixel with numerical simulation. The middle graph shows the absolute value of the difference between the numerical solution and the full-hypercube finite-gap approximation, multiplied by $10^{3}$. The difference at the left-hand side of the picture, of order $10^{-3}$, is very likely to be a numerical artifact. The graph below shows the absolute value of the difference between the full-hypercube finite-gap approximation and the approximation involving the relevant vertices only, multiplied by $10^{2}$; its magnitude is $O(10^{-2})$, a little higher than $\epsilon^{1/2}$ but, taking into account that the full hypercube contains $4^{6}=4096$ points, the agreement is sufficiently good (see Remark 11).}
\end{figure}

Suppose that, in some time interval $I$, only the $j^{th}$ unstable mode is visible (for example, in Figure~\ref{fig:3D_FG}, only the first mode $k_1$ is visible, and appears as an isolated peak in the center of the picture). Then, in the interval $I$, the pair of variables $\tilde w_j$, $\tilde w_{j+N}$ is close to $1/2$ and all other $\tilde w_k$ are either close to 0 or to 1. Then we have the following approximate formulas for the position $(X_{\mbox{max}},T_{\mbox{max}})$ of the isolated $j^{th}$ mode:
\begin{equation}
\label{eq:tmax}
T_{\mbox{max}} = T^{(1)}_j + {\cal M}_j \Delta T_j +\!\!\!\!\!\!\!\! \sum\limits_{\begin{array}{c}1\le k \le N \\ k\ne j \end{array}}\!\!\!\!\!\!\!\! {\cal M}_k \Delta_T(j,k),
\end{equation}
\beq\label{eq:xmax}
X_{\mbox{max}} = X^{(1)}_j + {\cal M}_j \Delta X_j. 
\eeq
$T^{(1)}_j,X^{(1)}_j$ would be the coordinates of the first appearance of the $j^{th}$ mode, and $\Delta T_j$ and $\Delta X_j$ would be the recurrence time and the $x$ shift of $j^{th}$ mode, if it did not interact with the other modes, and therefore are defined in (\ref{parameters_recurrence}). ${\cal M}_k$ indicates how many times the mode $k$ was visible before the time $T_{max}$:
\beq\label{number_appearances}
{\cal M}_k=\left\{\begin{array}{ll}  \mbox{nearest integer to} \ \ w_k, & k\ne j, \\
\lfloor w_j \rfloor & k=j, \end{array} \right.
\eeq
and
\beq
\Delta_T(j,k) = \frac{1}{\sigma_j}\log \left|\frac{\sin(\phi_j + \phi_k )}{\sin (\phi_j-\phi_k) }\right|
\eeq
is the time delay in the appearance of the $j^{th}$ mode due to its pairwise interaction with the $k^{th}$, $j\ne k$ mode. 
We see that that the presence of the other modes delays the appearance of the $j$-th mode (see (\ref{eq:tmax})), but does not affect the $x$-shift (see (\ref{eq:xmax})). We also see that the interaction of the unstable modes is pairwise.

\section{Acknowledgments} The work P. G. Grinevich was supported by the Russian Science Foundation grant 18-11-00316. P. M. Santini was partially supported by the University La Sapienza, grant 2017.\\
We acknowledge useful discussions with F. Calogero, C. Conti, E. DelRe, A. Degasperis, D. Pierangeli, M. Sommacal, and V. Zakharov.

\end{document}